\documentclass[12pt]{article}
\pdfoutput=1
\pdfoutput=1
\usepackage{amsmath,amssymb,amsfonts,bbold}
\usepackage{psfrag}
\usepackage{color}
\usepackage{epsfig}
\usepackage{graphicx}
\usepackage{epstopdf}
\usepackage{mathtools}
\usepackage{bbold}
\usepackage{hyperref}
\usepackage{tikz,pgfplots}
\usetikzlibrary{decorations.pathmorphing}
\tikzset{snake it/.style={decorate, decoration=snake}}
\usepackage{simpler-wick}
\def\spc{\hspace{1pt}}
\def\hhat{\widehat}
\usetikzlibrary{trees}
\usetikzlibrary{shapes.misc}
\usepackage{tkz-euclide}
\renewcommand{\tilde}[1]{\widetilde{#1}} 

\newcommand{\la}{\langle}
\newcommand{\ra}{\rangle}
\newcommand{\tr}{\textrm{tr}\,}
\newcommand{\bea}{\begin{eqnarray}}
\newcommand{\eea}{\end{eqnarray}}
\newcommand{\ea}{\end{eqnarray}}

\def\nn{\nonumber}

\def\be{\begin{eqnarray}}
\def\ee{\end{eqnarray}}
\def\hhR{h_R}
\def\hhL{h_L}
\def\Tr{{\rm Tr}}

\def\nspc{{\!\spc}}
\def\nsmpc{{\nspc\smpc}}
\def\bea{\begin{eqnarray}}
\def\eea{\end{eqnarray}}
\def\is{&\! = \! & }
\def\spc{\hspace{1pt}}
\def\smpc{\hspace{.5pt}}
\def\blueone{darkgray}
\def\bluetwo{blue}
\def\bluethree{red}

\def\zZ{{Z}}
\def\bfC{\mbox{\textbf{\textit C}}}
\renewcommand{\Large}{\large}
\renewcommand{\footnotesize}{\small}
\begin{document}

\begin{titlepage}

\setcounter{page}{1} \baselineskip=15.5pt \thispagestyle{empty}

\vfil

${}$
\vspace{1cm}

\begin{center}

\def\thefootnote{\fnsymbol{footnote}}
\begin{center}
{\Large \bf On the Quantum Information Content of a Hawking Pair}\\[5mm]
\end{center}

~\\[1cm]
Herman Verlinde
\\[0.3cm]

{\normalsize { \sl Physics Department,  
Princeton University, Princeton, NJ 08544, USA}} \\[3mm]

\end{center}

{\small  \noindent 
\begin{center} 
\textbf{Abstract}\\[4mm]
\parbox{15 cm}{
We introduce a new probe designed to keep track of the quantum information content of a Hawking pair as a function of the distance from the black hole horizon. We compute the entropy content of this Hawking pair probe via a semi-classical replica method that relies on free field Wick contractions and their leading order gravitational back reaction on the black hole horizon area. We find that the information transfer from the black hole state to the Hawking pair is triggered by a geometric transition that, somewhat surprisingly, takes place at a macroscopic distance from the horizon. We relate our computation to recent insights about the role of von Neumann algebras in holography.
}

\end{center} }
 \vspace{0.3cm}

\vfil
\begin{flushleft}

\today
\end{flushleft}

\end{titlepage}

\newpage
{\pagestyle{empty}
\tableofcontents}
\newpage

\def\calO{{b}}
\def\be{\begin{equation}}
\def\ee{\end{equation}}

\renewcommand{\Large}{\large}
\def\nn{\nonumber}


\def\tr{{\rm tr}}
\def\li{|\spc}
\def\ri{|\spc}
\def\stau{{\mbox{\footnotesize$\tau$}}}
\def\mini{\raisebox{1pt}{\tiny$\smpc-$}}
\def\plus{\mbox{\tiny$+$}}
\def\star{{}}

\addtolength{\abovedisplayskip}{.75mm}
\addtolength{\belowdisplayskip}{.75mm}
\addtolength{\parskip}{.75mm}
\setcounter{page}{1}
\section{{{Introduction}}}
\vspace{-1mm}

The Page curve is one of few quantitative diagnostics of the black hole information paradox. While usually stated in terms of the time evolution of the entropy of black hole radiation, it can also be viewed as a prediction about the incremental amount of quantum information that each individual Hawking pair must extract from the black hole to preserve unitarity: a Hawking pair emission that reduces the black hole mass by $\Delta M$ must carry along with it $\Delta S = \beta \Delta M$ amount of quantum information \cite{Hawking}-\cite{jacobson}. The outstanding challenge is~to find out how this information transfer takes place \cite{AMPS}.

In this paper we will ask a simpler question: where and when does the information transfer from the black hole to the Hawking pair take place? To investigate this question and make it quantitative, we propose a new probe that keeps track of the quantum information content of a Hawking pair as a function of its location. Virtual Hawking pairs in the vicinity of the horizon are true vacuum fluctuations \cite{Unruh}, and for this reason, they are in a uniquely entangled state that is not expected to contain any detailed information about the black hole micro-state. On the other hand, unitarity prescribes that far away Hawking pairs should extract a non-zero amount of quantum information from the black hole. Our goal is to give a quantitative description of the transition \cite{Giddings} between the two regimes.

We will work within the context of AdS/CFT at large $N$ and assume that the Hawking particles are heavy in AdS units. In this regime, familiar semi-classical methods apply. We would like to use these methods to identify where the predictions from holography differ from those of QFT in curved spacetime. For this purpose, it will be useful to formulate parts of our set up in the language of operator algebras. Our calculations, however, will be based on familiar elements of semi-classical gravity, CFT, and the holographic dictionary. 

To describe the black hole of mass $M$, we introduce a general state of the CFT that is maximally entangled with some auxiliary reference system
\bea
\label{micropure}
|\Psi \rangle \is  \sum_{n} \,  \spc |E_n \rangle\spc |\psi_n\rangle_{\rm aux}\spc .
\eea
Here the sum  is restricted to a small energy window of width $\Delta E$ centered around some average energy $E\!=\!M$. The reference system is used as a way to codify the quantum information of the black hole state. We will call $\Psi$ the micro-canonical CFT state.

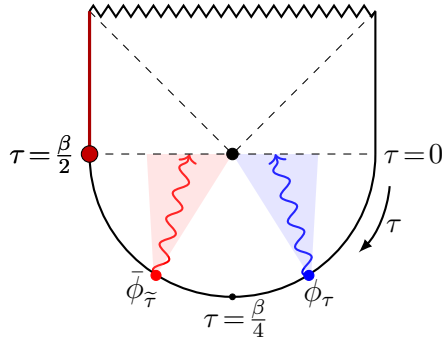
\begin{figure}[t]
\begin{center}
\begin{tikzpicture}[scale=.95]
\path [draw={rgb:red,5; black,0}, ->, thick,snake it] (-1.05,-1.75) -- (-.6,0);
\path [draw={rgb:blue,3; black,0}, ->, thick,snake it] (1.1,-1.75) -- (.6,0);
\path[fill={rgb:red,2; white,3},opacity=.25]  (-1.1,-1.75) -- (-1.2,0) -- (-.0,0) --cycle ;
\path[fill={rgb:blue,2; white,3},opacity=.25]  (1.1,-1.75) -- (1.2,0) -- (.0,0) --cycle ;
\draw[dashed] (-2,0) -- (2,0);
\draw[thick] (-2,0) -- (-2,2);
\draw[thick] (2,0) -- (2,2);
\draw[dashed] (-2,2) -- (0,0);
\draw[dashed] (2,2) -- (0,0);
\draw[fill=black] (0,0) circle (0.08);
\draw[thick,decoration = {zigzag,segment length = 2mm, amplitude = 0.75mm},decorate] (-2,2)--(2,2);
\draw[thick] (-2,0) arc (-180:0:2);
\draw[very thick,color = {rgb:red,10; black,5}] (-2,0) -- (-2,2);
\draw[fill={rgb:red,10; black,3}] (-2,0) circle (0.115);
\draw (1.22,-1.95) node {$\phi_\tau$};
\draw (-1.27,-1.9) node {$\bar{\phi}^{\star}_{\tilde\tau}$};
\draw (-2.6,-0) node {\small $\tau\! = \! \textstyle \frac \beta 2\ $};
\draw (.05,-2.3) node {\small $\tau \! =\!\textstyle \frac \beta 4$};
\draw (-2.6,-0) node {\small $\tau\! = \! \textstyle \frac \beta 2\ $};
\draw (2.4,-0) node {\small $\ \ \tau \! = \! 0$};
\draw[thick,-latex] (2.2,-0.45) arc (-5:-45:1.5);
\draw (2,-1) node[right] {\small $\tau$};
\draw[fill= black] (0,-2) circle (0.04);
\draw[color={rgb:red,10; black,0}, fill={rgb:red,10; black,0}] (-1.07,-1.7) circle (0.07);
\draw[color={rgb:blue,10; black,0}, fill={rgb:blue,10; black,0}] (1.07,0-1.7) circle (0.07);
\end{tikzpicture}
\end{center}
\vspace{-6mm}
\caption{The hybrid space-time with a euclidean black hole transitioning to a lorentzian black hole. A Hawking pair is created at points on the past boundary, placed on opposite sides of the horizon. 
The red dot and line on the left denote an end-of-the-world brane. }
\vspace{-2mm}
\end{figure}

We are interested in the properties of  the CFT state $\Psi$ as seen by a bulk observer. Let ${\cal A}$ denote the algebra of CFT operators that via the HKLL prescription are identified with the operator algebra of a free bulk quantum field theory defined on the black hole exterior \cite{BDHM}-\cite{FL}. The state $\Psi$ looks like a thermo-field double (the purification of a thermal state) from the point of view of ${\cal A}$. A bulk observer will thus experience $\Psi$ as the Hawking state \cite{PR2}-\cite{wittenx} of the black hole with mass $E$. The eigenstate thermalization hypothesis implies that the bulk time flow generated by the modular hamiltonian $H_\Psi$ is physically indistinguishable from the boundary time flow generated by the microscopic CFT Hamiltonian $H_{\rm CFT}$.

Let $\phi_0 \in {\cal A}$ denote a local quantum field operator, or single trace CFT operator, acting on the spatial AdS boundary at lorentzian time $t=0$. The operator
${\phi}_{\nspc{\tau}} = e^{-\tau H_\Psi} \phi_0 e^{\tau H_\Psi}$ then acts at some euclidean time $\tau$ along the past AdS boundary of the euclidean black hole geometry \cite{MaldacenaTFD,Kourkoulou}. The $\Psi$ expectation value of two such local operators ${\phi}_{\nspc{\tau_1}}$ and $\bar\phi_{\nspc{\tau_2}}$ coincides with euclidean thermal CFT two-point function 
\bea 
\label{thermaltwo}
\la\Psi |  \bar{\phi}_{{\tau_2}} \spc{\phi}_{{\tau_1}} \nspc |\Psi\ra\! \is \!
  \frac 1 {Z_\beta}\spc {\rm Tr}\bigl(\nspc e^{-\beta H_{\Psi}}  \bar{\phi}_{\!{\tau_2}} \spc {\phi}_{\nspc{\tau_1}}\nspc \bigr)\equiv 
G_{\beta}(\tau_{12})
\eea
at inverse temperature $\beta=\beta_E$. Via the holographic map, the two point function $G_{\beta}(\tau_{12})$ represents the boundary-to-boundary propagator of the free bulk particle created by the $\phi$ field operator. This two-point-function will play a central role in the following.

In an upcoming paper \cite{toappearsoon} we study the properties of the state $\Phi_1 = \phi_\tau \Psi$ with one extra particle obtained by acting with a local operator $\phi_\tau$ on the Hawking state $\Psi$ and show how it can be used to probe the bulk black hole space-time both outside and inside of the horizon. In this paper, we will instead study the properties a state $\Phi$ with an extra Hawking pair at a specified location. This Hawking pair state $\Phi$ is obtained by acting on the state~$\Psi$ with the product of a local CFT operator $\phi_\tau$ and its mirror operator $\tilde\phi_\tau$.
The mirror operator $\,\tilde{\!\phi_\tau\!}\,$ is designed such that it amounts to inserting a conjugate operator $\bar{\phi}_{\spc\tilde\tau}$  behind the horizon at the mirror location
\bea
\label{ttilde}
\tilde{\tau} \is \beta/2 - \tau,
\eea
as indicated in figure 1. The Hawking pair state thus takes the form 
\bea
\label{phidef}
|\Phi \ra\! \is 
\frac{1}{\mbox{$\textstyle \sqrt{N\ \ }{\!\!\!\!\!\spc}_{\beta,\tau}$}\!} \;
 {\phi}_\tau \spc \bar{\phi}_{\spc\tilde\tau}\spc | \Psi \ra.
\eea
The normalization factor $N_{\beta,\tau}$ is given by the four-point function
\bea
\label{fourpoint}
N_{\beta,\tau} = \spc {Z_{\beta,\tau}}/Z_\beta \is
  \la\Psi|\spc  {\phi}_{-\tilde\tau } \spc \bar{\phi}_{-\tau} \spc{\phi}_{\tau} \spc \bar{\phi}_{\tilde\tau}\spc |\Psi\ra\ \
\eea

The local CFT operators $\phi_\tau$ and $\bar{\phi}_{\spc\tilde\tau}$ create bulk excitations that propagate to the time slice $t=0$, where they appear as free field excitations smeared via the boundary-to-bulk propagator. For heavy particles compared to the AdS scale, the smearing function localizes around the classical trajectory and the boundary location $\tau$ of $\phi_\tau$ provides a reliable proxy for the radial location of the corresponding Hawking particle at $t=0$. We wish to compute the quantum information content of this Hawking pair state as a function of the location~$\tau$. 
The main assumption that goes into our analysis is that correlation functions of $\phi_\tau$ obey the rules on free field theory and that gravity can be treated semi-classically.

\smallskip

We now state our main result. Let $\rho_\Phi$ and $\rho_\Psi$ denote the density matrices associated with the states $\Phi$ and $\Psi$. Our goal is to compute the difference 
\bea
\label{deltas}
\Delta S \is 
S(\rho_{\Phi\nspc})\nspc -\nspc S(\rho_\Psi)
\eea
of the von Neumann entropies $S(\rho_{\Phi}) = -\Tr \rho_\Phi \log \rho_{\Phi}$ and $S(\rho_{\Psi}) = -\Tr\rho_\Psi \log \rho_{\Psi}.$
We will interpret the entropy difference \eqref{deltas} as a measure of the quantum information content associated with the Hawking pair. 
In QFT in a fixed curved space-time, this entropy difference \eqref{deltas} is expected to be positive and equal to the extra entanglement entropy across the horizon created by the presence of the extra Hawking pair. In a holographic setting, however, we expect the entropy difference to receive a negative gravitational contribution proportional to the decrease in the horizon area due to the emission of the Hawking pair. As we will see, in the semi-classical limit of interest, the negative gravitational contribution is parametrically larger than the positive QFT contribution.

It will be practical to introduce the coordinate differences
 \bea
 x\spc  =\spc 2\tau, \qquad \tilde{x} \is   \tau - \tilde{\tau}\spc =\spc 2\tau\nspc - \nspc \beta/2.
 \eea
Here $\tau$ runs from $0$ to $\beta/4$. We also introduce two coordinate regions $I$ and $I\nspc I$ 
separated by a transition point $\tau_c = \beta/8$, situated midway between the horizon location $\tau = \beta/4$ and the asymptotic AdS boundary located at $\tau=0$
\bea
 I \; :\ \, \mbox{near horizon region}\is \; \bigl\{ x> \tilde x \spc \bigr\} \ = \, \{ \beta/4 > \tau> \beta/8 \}
 \quad \nn \\[-2mm]\\[-2mm]\nn
 I\nspc I \, : \spc \ \ \;   \ \, \mbox{far  away region}\ \  \is \; \bigl\{ x< \tilde x \spc \bigr\} \ = \ \{\,0 \, < \tau < \beta/8 \spc \}\, .\quad
\eea 
Our final result for the leading order semiclassical contribution to the entropy difference $\Delta S$ between the state with and without the extra Hawking pair can then be expressed in terms of 
the thermal two-point function \eqref{thermaltwo} as follows
\bea
\label{deltasfin}
 \Delta S \is   \left\{\! \begin{array}{cc} 0 \qquad\qquad  & \quad  \raisebox{-1pt}{$I$}\ \\[2mm]
-n  \frac{\raisebox{2pt}{$\partial\ $}}{\raisebox{-3.5pt}{$\partial n$}} \log\biggl( \frac{\raisebox{3pt}{$G_{n\beta}(\spc\tilde{x}\spc
)^2$}}{\raisebox{-4pt}{$G_{n\beta}(\spc x\spc)^2$}}\biggr)_{\bigl|n=1} & \quad \raisebox{-3pt}{$I \nspc I$}\  \end{array}\right. \ \ 
 \eea
The entropy difference vanishes in the near horizon region $I$, indicating that the Hawking pair has not yet extracted any entropy from the black hole. After passing the transition point $\tau=\tau_c$, the Hawking pair starts extracting quantum information from the black hole. As we will see, the above formula for $\Delta S$ represents a monotonically decreasing function in region $II$, reaching a maximum negative value $\Delta S(\tau=0)= -4 h$, with $h$ the conformal  dimension of $\phi_\tau$, as $\tau$ approaches the AdS-boundary at $\tau=0$.

The non-zero contribution to $\Delta S$ in \eqref{deltasfin} represents the black hole entropy reduction due to the gravitational backreaction of the physical Hawking pair. To understand the physical origin of the transition at $\tau_c=
\beta/8$, it is useful to consider the four point function \eqref{fourpoint}.
We can think of this four point function $Z_{\beta,\tau}$ as the partition function of the Hawking pair.
At large $N$,  $Z_{\beta,\tau}$ can be evaluated by means of Wick's theorem. There are two possible Wick contractions. 
 In the semi-cassical regime, ${Z}_{\beta,\tau}$ exhibits a sharp transition at $\tau_c = \beta/8$, at which one~of the two Wick contractions becomes dominant over the other, depending on  which boundary-to-boundary geodesics is the shortest.  
 \bea
\label{twowick}
 ~~{I \, : }~~~~\raisebox{2.5pt}
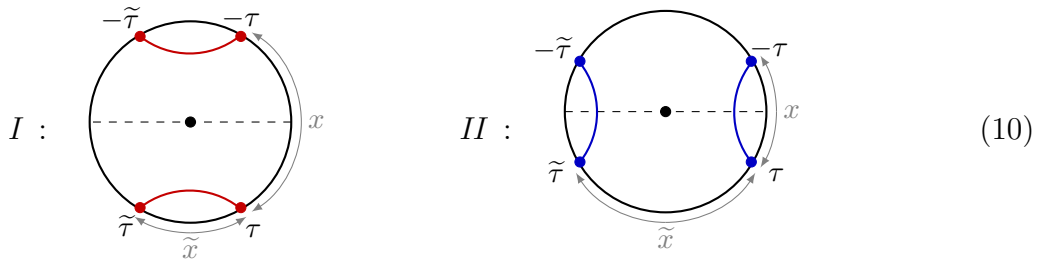
{\begin{tikzpicture}[scale=.67, baseline={([yshift=0cm]current bounding box.center)}]);
\draw[thick] (4,0) arc (0:360:2);
\draw[color={rgb:red,10; black,3}, fill={rgb:red,10; black,3}]   (1,-1.7) circle (0.1); 
\draw[color={rgb:red,10; black,3}, fill={rgb:red,10; black,3}]   (3,-1.7) circle (0.1); 
\draw[color={rgb:red,10; black,3}, fill={rgb:red,10; black,3}]   (1,1.7) circle (0.1); 
\draw[color={rgb:red,10; black,3}, fill={rgb:red,10; black,3}]   (3,1.7) circle (0.1); 
\draw[thick, color={rgb:red,10; black,3}] (1,-1.7) arc (90+38:90-38:1.6);
\draw[thick, color={rgb:red,10; black,3}] (1,1.7) arc (270-38:270+38:1.6);
\draw[thin,-latex,color=gray] (4.2,0) arc (0:-58:2.1);
\draw[thin,-latex,color=gray] (4.2,0) arc (0:58:2.1);
\draw[thin,-latex, color=gray] (2,-2.2) arc (270:238:2.1);
\draw[thin,-latex, color=gray] (2,-2.2) arc (270:302:2.1);
\draw (4.5,-0) node {\mbox{\textcolor{gray}{ \footnotesize ${x}$}}};
\draw (2,-2.5) node {\mbox{\textcolor{gray}{ \footnotesize $\tilde{x}$}}};
\draw (3.25,-2.15) node {\small $\tau$};
\draw (.7,-2.05) node {\small $\spc\tilde\tau$} ;
\draw (3.1,2.05) node {\small ${\!\!-\tau}$};
\draw (.7,2.05) node {\small ${\!\!-\tilde\tau}$} ;
\draw[dashed] (4,0) -- (0,0);
\draw[fill=black] (2,0) circle (0.1);
\end{tikzpicture}} 
\qquad \qquad {I\nspc I\,  : \ \ }\raisebox{2pt}{\begin{tikzpicture}[scale=.67, rotate=90, baseline={([yshift=0cm]current bounding box.center)}]);
\draw[thick] (4,0) arc (0:360:2);
\draw[color={rgb:blue,10; black,3}, fill={rgb:blue,10; black,3}]   (1,-1.7) circle (0.1); 
\draw[color={rgb:blue,10; black,3}, fill={rgb:blue,10; black,3}]   (3,-1.7) circle (0.1); 
\draw[color={rgb:blue,10; black,3}, fill={rgb:blue,10; black,3}]   (1,1.7) circle (0.1); 
\draw[color={rgb:blue,10; black,3}, fill={rgb:blue,10; black,3}]   (3,1.7) circle (0.1); 
\draw[dashed] (2,2) -- (2,-2);
\draw[fill=black] (2,0) circle (0.1);
\draw[thin,-latex,color=gray] (-.2,0) arc (180:122:2.1);
\draw[thin,-latex,color=gray] (-.2,0) arc (180:238:2.1);
\draw[thin,-latex, color=gray] (2,-2.2) arc (270:238:2.1);
\draw[thin,-latex, color=gray] (2,-2.2) arc (270:302:2.1);
\draw (-.5,-0) node {\mbox{\textcolor{gray}{ \footnotesize $\tilde{x}$}}};
\draw (2,-2.5) node {\mbox{\textcolor{gray}{ \footnotesize ${x}$}}};
\draw[thick, color={rgb:blue,10; black,3}] (1,-1.7) arc (90+38:90-38:1.6);
\draw[thick, color={rgb:blue,10; black,3}] (1,1.7) arc (270-38:270+38:1.6);
\draw (.75,-2.2) node {\small $\tau$};
\draw (.8,2.2) node {\small ${\spc\tilde\tau}$} ;
\draw (3.25,-2.15) node {\small ${\!\!-\tau}$};
\draw (3.25,2.15) node {\small ${\!\!-\tilde\tau}$} ;
\end{tikzpicture}}
\eea
 Taking this into account, we find that 
\bea
\label{twocases}
\ Z_{\beta,\tau}/Z_\beta 
\is  
\spc G_{\beta} (\spc y \spc)^2\qquad\, \quad  y = \min(x,\tilde{x}) 
\eea
In the near horizon region $I$, the field operator $\phi_\tau$ is closest to its mirror partner $\bar{\phi}_{\tilde{\tau}}$ behind the horizon and the corresponding Wick contraction, indicated in the left figure below, dominates. The Hawking pair in this regime is still virtual and does not backreact on the horizon area. The black hole entropy remains unchanged in this case.
In the far away region $I\nspc I$, on the other hand, the Hawking particle $\phi_\tau$ is closest to its hermitian conjugate $\bar{\phi}_{-\tau}$ and the corresponding Wick contraction dominates. Since this contraction crosses the $t=0$ time slice, the Hawking pair is now physical and does backreact on the horizon area. We will compute the resulting gravitational contribution in several ways.

This paper is organized as follows. In section 2 we give a definition of the state $\Phi$ of the Hawking pair in terms of operator algebras. This somewhat formal definition will be useful for connecting our results with the recent understanding of the role of type II algebras in semi-classical holography. In section 3 we introduce the partition function of the Hawking pair and review the microscopic interpretation of the holographic Wick contractions. In section 4 we compute the entropy difference $\Delta S$ via three methods: via a thermodynamic argument, by means of  a replica calculation, and as a limit of exact results in JT gravity. The three methods all give the same answer \eqref{deltasfin}.
These calculations closely follow the earlier work \cite{GLTV} on partially entangled states in JT gravity. In section 5 we compute $\Delta S$ by making use of the recent insights in the relationship between the semi-classical limit of holography and type II and III von Neumann algebras \cite{LL1}-\cite{wittenetal}. We end with some comments on the physical interpretation and implications of our study. Some technical parts of our discussion are referred to the Appendix.

\medskip

\section{{{Quantum state of a Hawking pair}}}
\vspace{-1mm}

In this section we introduce the quantum state of a Hawking pair. Our construction is general, but for concreteness we will specialize to the context of an AdS black hole of mass $M$ dual to a state $\Psi$ of the form \eqref{micropure} in a holographic CFT. For later discussion, it will be useful to give some definitions in the language of modular theory.\footnote{For a review of the basics of modular theory, see \cite{witten18}. Readers that are less formally inclined can simply skip section 2.1 and the second paragraph of section 2.3.}

\subsection{{Hawking state}}
\vspace{-1mm}

Let   ${\cal H}$ denote the Hilbert space of the CFT and the auxiliary system and ${\cal A}$ the algebra of light HKLL CFT operators associated to the black hole exterior.
In the large $N$ limit, the bulk QFT is non-interacting and gravity is semi-classical. 
The microcanonical state $\Psi$ given in \eqref{micropure} defines a cyclic and separating vector in~${\cal H}$. 
This allows us to introduce  the anti-linear Tomita operator $S_\Psi$ defined by the property 
$S_\Psi {\bf a} |\Psi\rangle ={\bf a}^\dag |\Psi\rangle$  
for any ${\bf a} \in {\cal A}$. The modular operator $\Delta_\Psi$ and Hamiltonian $h_\Psi$ associated to ${\Psi}$ are then defined via 
\bea 
\label{modularo}
\Delta_\Psi\! \is S_\Psi^\dag S_\Psi = \, e^{- \beta h_\Psi}.\
\eea 
In a thermalizing CFT, the modular flow generated by the unitary operator $\Delta_\Psi^{it/\beta} = e^{-i t h_\Psi}$
coincides with the time flow generated by the microscopic CFT Hamiltonian. Via this flow, we can associate to a local single trace operator $\phi_0 \in {\cal A}$ defined at the timeslice $t=0$ a one-parameter family of local single trace operators $\phi_\tau$ along the euclidean time direction 
\bea
\label{modflow}
{\phi}_\tau \is e^{-\tau h_\Psi}\! {\phi}_0\spc e^{\tau h_\Psi}.
\eea 
Correlation functions satisfy the KMS condition
\bea
\label{kms}
\la \Psi | {\phi}_\tau {\bf a} |\Psi\ra \is \la \Psi | {\bf a}\spc \Delta_\Psi {\phi}_\tau |\Psi\ra \spc =\spc \la \Psi| {\bf a} \spc {\phi}_{\tau + \beta}|\Psi\ra.
\eea
for any operators ${\bf a} \in {\cal A}$. Hence we can view $\tau$ as a coordinate along the thermal circle.  In holographic terms, $\Psi$  looks like the Hawking state from the perspective of the bulk QFT. 

\subsection{{Hawking state with one extra particle}}
\vspace{-1mm}

In the upcoming paper \cite{toappearsoon} we study the properties of the state $\Phi_1 = \phi_\tau \Psi$ obtained by acting with a light local operator $\phi_\tau$ on a heavy state $\Psi$ in a holographic CFT. $\Phi_1$ represents a single particle state in the bulk, created at the past boundary of the euclidean black hole. The time $\tau$ along the thermal circle is mapped via radial quantization to the radial distance $|z| = e^\tau$ between $\phi_\tau \equiv \phi(z)$ and the heavy operator $\Psi(0)$. Holography predicts that at the special radial distance  $\tau=\beta/4$, the wave function of the bulk particle transitions from being (mostly) outside to being (mostly) inside the black hole horizon. In pictures,
\bea
~~\raisebox{0cm}{$\phi_\tau |{\Psi} \rangle\spc = \, e^{-\tau h_\Psi} \phi_0 |\Psi\ra \, = \ $}~\raisebox{-1.25cm}{$\begin{tikzpicture}[scale=.66]
\draw[dashed] (-2,0) -- (2,0);
\draw[thick] (-2,0) arc (-180:0:2);
\draw[thin, -latex] (-2,0) arc (-180:0:1.99);
\draw[fill=black] (0,0) circle (0.08);
\draw[fill={rgb:red,10; black,3}] (-2,0) circle (0.08);
\path [draw={rgb:blue,5; black,2}, ->, thick,snake it] (1.1,-1.75) -- (.6,0);
\path[fill={rgb:blue,2; white,3},opacity=.35]  (1.1,-1.75) -- (1.3,0) -- (-0,0) --cycle ;
\draw[color={rgb:blue,10; black,3}, fill={rgb:blue,10; black,0}] (1,0-1.76) circle (0.08);
\draw[fill=black](0,-2) circle (0.06);
\draw (1.3,-2.1) node {\small $\tau$};
\draw (.05,-2.4) node {\footnotesize $\beta/4$};
\end{tikzpicture}$} \nn \\[-3mm]\\[1mm]
\raisebox{0cm}{$\bar\phi_{\tilde{\tau}}|{\Psi} \rangle \, =\, e^{\tau h_\Psi} \bar\phi_{\beta/2} |\Psi\ra \, = \ $}~\raisebox{-1.25cm}{$\begin{tikzpicture}[scale=.66]
\draw[dashed] (-2,0) -- (2,0);
\draw[thick] (-2,0) arc (-180:0:2);
\draw[thin, -latex] (-2,0) arc (-180:0:1.99);
\draw[fill=black] (0,0) circle (0.08);
\draw[fill={rgb:red,10; black,3}] (-2,0) circle (0.08);
\path [draw={rgb:red,5; black,0}, ->, thick,snake it] (-1.05,-1.75) -- (-.6,0);
\path[fill={rgb:red,2; white,3},opacity=.25]  (-1.1,-1.75) -- (-1.3,0) -- (0,0) --cycle ;
\draw[color={rgb:red,10; black,3}, fill={rgb:red,10; black,0}] (-1,-1.76) circle (0.08);
\draw (-1.3,-2.1) node {\small ${\tilde\tau}$};
\draw[fill=black](0,-2) circle (0.06);
\draw (.05,-2.4) node {\footnotesize $\beta /4$};
\end{tikzpicture}$} \nn
\eea
with $\tilde\tau = \beta/2-\tau$. Both  are well-defined normalized states over the range $0\leq \tau \leq \beta/2$. As shown in \cite{toappearsoon}, scaling and universality of the OPE relation in a holographic CFT imply that as $\phi_\tau$ approaches $\Psi(0)$, the state $\Phi_1=\phi_\tau \Psi$ undergoes a transition from having larger energy to having smaller energy than $\Psi$. This transition occurs at the mid-point $\tau=\beta/4$ and marks the place where~the bulk particle created by $\phi_\tau$ passes through the horizon.

We will work in the semi-classical regime where $G_N$ is small and the bulk particle created by $\phi_\tau$ is heavy compared to the AdS-scale. In this regime, particles propagate along classical geodesics. The classical trajectory associated with the two point~function 
\bea
\label{saddle}
 \raisebox{0cm}{$\la \Psi| \bar\phi_{-\tau} \phi_\tau |{\Psi} \rangle \, = \, G_\beta(2\tau) 
\, = \spc $}~~\raisebox{-1.25cm}{$\begin{tikzpicture}[scale=.66]
\draw[thick,  color={rgb:blue,10; black,3}] (-2.7,1.5) arc (180-38:180+38:2.45);
\draw[dashed] (-6,0) -- (-2.1,0);
\draw[thick] (-2,0) arc (360:0:2);
\draw[fill=black] (-4,0) circle (0.08);
\draw[color={rgb:blue,10; black,3}, fill={rgb:blue,10; black,3}] (-5.9+3.2,1.5) circle (0.06);
\draw[color={rgb:blue,10; black,3}, fill={rgb:blue,10; black,3}] (-5.9+3.2,-1.5) circle (0.06);
\draw[xscale=1.15,rotate=90,color={rgb:blue,10; black,3}, fill={rgb:blue,10; black,3}] (0,2.785) circle (0.05);
\draw[xscale=1, rotate=90, fill={rgb:red,10; black,3}] (0,6) circle (0.08);
\draw (-2.75,-1.65) node[right] {\small $\tau$};
\draw (-2.78,1.65) node[right] {\small $-\tau$};
\draw (-3.35,.28) node[right] {\small $r$};
\end{tikzpicture}$}\ \ \ 
\eea
provides an identification between the euclidean time $\tau$ and the radial location $r$ along the $t=0$ time slice. We can make this relation more explicit for the case of 3D AdS gravity. The euclidean BTZ black hole geometry corresponding to the state $\Psi$ takes the form 
\bea
ds^2 = \Bigl(\frac{2\pi}{\beta}\Bigr)^{\nspc 2}\!\tan^2 \Bigl(\frac{2\pi r}{\beta}\Bigr)\spc dt^2 + \frac{d r^2+d{\phi}^2}{\Bigl(\nspc\mbox{\large $\frac{\beta}{2\pi}$}\nspc\Bigr)^2\! \cos^2 \Bigl(\mbox{\large $\frac{2\pi r}{\beta}$}\Bigr)} \label{btz}
\eea
 Here $r$ runs from $-\beta/4$ to $\beta/2$, with the horizon located at $r=0$.
As depicted in figure \eqref{saddle}, a bulk point at radial location $r$ on the $t=0$ time~slice is connected through a euclidean geodesic to two boundary points at $
\pm\tau\nspc =\nspc \pm (r\!-\!\beta/4).$
This geodesic is the worldline of a particle at rest at $t=\!0$. It extends into the black hole~interior when $\tau>\beta/4$. As shown in \cite{toappearsoon}, the expectation value of the CFT Hamiltonian in the the state $\Phi_1$ precisely matches with the Schwarzschild energy of a $\phi$ particle at rest at the corresponding radial location~$r$.

\subsection{{Hawking pair state}}
\vspace{-1mm}

We now introduce the Hawking pair state $\Phi$ as the special state obtained by acting with two QFT operators on opposite sides of the black hole horizon, thus creating the Hawking particle and its partner.
The relative location of the pair is fixed by the condition that the combined pair has zero energy: if the exterior particle has positive energy $\Delta E$, the interior particle should have negative energy $-\Delta E$. Concretely, we will require that the energy of the Hawking pair state $\Phi$,
defined as the expectation value of the CFT Hamiltonian $H_{\rm CFT}$,  is equal to the energy of the Hawking state $\Psi$ 
\bea
\label{econd}
\la \Phi | H_{\rm CFT}  |\Phi\ra \is \la \Psi |H_{\rm CFT} |\Psi\ra \, = \, E.
\eea

To write the Hawking pair state we will makes use of the PR mirror map \cite{PR2}.  The mirror map is an anti-linear isometry $J_\Psi = J_\Psi^{-1}$ defined through the polar decomposition  $S_\Psi =  J_\Psi \Delta_\Psi^{1/2} = \Delta_\Psi^{-1/2} J_\Psi$ of the Tomita operator. It associates to the local CFT operator $\phi_\tau$  a mirror operator 
 \bea
 \widetilde{\phi_\tau\!} \is J_\Psi \phi_\tau J_\Psi.  
 \eea
 Physically, $J_\Psi$ interchanges the outside Hawking particle with its interior partner.  To make this explicit, we combine the above definition of $J_\Psi$ with the definition \eqref{modflow} of the modular time flow to rewrite the mirror operator as
\bea
 \widetilde {\phi_{\tau}\!}   \is
 S_\Psi \spc \phi_{\tilde{\tau}}\spc S_\Psi,\  \ \ \  \ \tilde \tau \spc \equiv  \spc \beta/2 - \tau
\eea
$ \tilde \tau$ is the mirror location dual to $\tau$, as shown in figure 1.

\setcounter{equation}{3}

We now define the Hawking pair state as in equation (4) repeated here
\bea
\label{hpair}
 |\smpc\Phi\smpc\ra \is 
\frac{1}{\mbox{$\textstyle \sqrt{N\ \ }{\!\!\!\!\!\spc}_{\beta,\tau}$}\!} \; \tilde{\phi_\tau\!} \; {\phi_{\tau}}\spc | \Psi \ra \, 
\, \equiv \, 
\frac{1}{\mbox{$\textstyle \sqrt{N\ \ }{\!\!\!\!\!\spc}_{\beta,\tau}$}\!} \; \phi_{\tau}\spc  \bar{\phi}_{\tilde{\tau}}| \Psi \ra \,  
\eea
with $N_{\beta,\tau}$ the four point function given in equation~\eqref{fourpoint}.  Here is a pictorial representation of the Hawking pair state
\setcounter{equation}{20}
\bea
|{\Phi} \ra  \, \is  ~~~\raisebox{-.5cm}{$\begin{tikzpicture}[scale=.68]
\draw[dashed] (-2,0) -- (2,0);
\draw[thick] (-2,0) arc (-180:0:2);
\draw[thin, -latex] (-2,0) arc (-180:0:1.99);
\draw[fill=black] (0,0) circle (0.05);
\draw[fill={rgb:red,10; black,3}] (-2,0) circle (0.08);
\path [draw={rgb:red,5; black,0}, ->, thick,snake it] (-1.05,-1.75) -- (-.6,0);
\path [draw={rgb:blue,5; black,2}, ->, thick,snake it] (1.1,-1.75) -- (.6,0);
\path[fill={rgb:red,2; white,3},opacity=.25]  (-1.1,-1.75) -- (-1.3,0) -- (0,0) --cycle ;
\path[fill={rgb:blue,2; white,3},opacity=.35]  (1.1,-1.75) -- (1.3,0) -- (0,0) --cycle ;
\draw[color={rgb:black,10; black,3}, fill={rgb:red,10; black,0}] (-1,-1.76) circle (0.08);
\draw[color={rgb:black,10; black,3}, fill={rgb:blue,10; black,0}] (1,0-1.76) circle (0.08);
\end{tikzpicture}$}_{\strut}
\eea
In bulk language, it defines the wave function on the $t=0$ slice obtained by performing the QFT path integral over the lower half of the euclidean black hole geometry with the insertion of two local operators on the past boundary. So it looks like it contains two extra (virtual or real) particles relative to the pure Hawking state $\Psi$. 

A better way to interpret $\Phi$ is that it represents the {\it component} of the Hawking state that contains a given Hawking pair. Indeed, $\Phi$ is not orthogonal to $\Psi$, but has finite overlap
\bea
\la \Phi|\Psi\ra 
\is G_\beta(\tau\!-\!\tilde\tau)~=~\raisebox{-1.25cm}{$\begin{tikzpicture}[scale=-.66, rotate = 180]
\draw[thick,  color={rgb:red,10; black,3}] (-2.55,1.45) arc (270+37:270-37:2.4);
\draw[dashed] (-6,0) -- (-2.1,0);
\draw[thick] (-2,0) arc (360:0:2);
\draw[fill=black] (-4,0) circle (0.08);
\draw[color={rgb:red,10; black,3}, fill={rgb:red,10; black,3}] (-5.8+3.2,1.45) circle (0.06);
\draw[color={rgb:red,10; black,3}, fill={rgb:red,10; black,3}] (-5.85+.5,1.45) circle (0.06);
\draw[xscale=1, rotate=90, fill={rgb:red,10; black,3}] (0,6) circle (0.08);
\draw (-6.15,1.65) node[right] {\small $\tilde\tau$};
\draw (-2.65,1.6) node[right] {\small $\tau$};
\end{tikzpicture}$}
\eea
The collection of all Hawking pair states of the form \eqref{hpair} define a natural basis of the space of two particle states that have overlap with $\Psi$  and are subject to the energy condition \eqref{econd}.
Note that all states of the form $\Phi$ are symmetric under modular conjugation by $J_\Psi$. 

We will treat the euclidean times $\tau$ and $\tilde\tau$ as a proxy for the radial location  $r$ of the Hawking pair on the $t=0$ slice. The two notions are semi-classically related via the two particle version of the saddle point argument in equation \eqref{saddle}.  In general, the relation between $\tau$ and $r$ involves a non-trivial smearing with bulk-to-boundary propagators. 

 As noted in the introduction, the four-point function ${Z}_{\beta,\tau}$ is a sum of two Wick contractions. This tells us that the probability distribution associated to $\Phi$ is concentrated around two classical trajectories indicated in equation \eqref{twowick}. In the first (red) contraction, the Hawking pair annihilates before it reaches the $t=0$ slice. In the second (blue)  contraction, the Hawking pair emerge at $t=0$ as two physical particles. 
The presence of the two terms indicates
that a bulk observer sees the Hawking pair as an entangled sum $|\psi_{\rm pair}\ra = \alpha_0 |0\ra |0\ra + \alpha_1 |1\ra |1\ra,$ where $|0\ra$ and $|1\ra$ mark whether each particle is present or not.  As we will see, the probabilities $p_i=|\alpha_i|^2$  depend on the location $\tau$. In the near horizon region $I$ the vacuum probability $p_0$ dominates, whereas in the far away regime $I\nspc I$, the probability $p_1$ that both particles are present quickly approaches unity. 

\section{{{Partition Function of a Hawking pair}}}

\vspace{-1.5mm}

Assuming the algebra ${\cal A}$ of external HKLL operators comes equipped with a trace, we can associate to the state $\Psi$ and algebra ${\cal A}$ a reduced density matrix $\rho_\Psi$ defined via the property that $\la \Psi| {\bf a} |\Psi\ra = {\rm Tr}(\rho_\Psi {\bf a})$ for all ${\bf a} \in {\cal A}$. We can write this density matrix in terms of a  corresponding modular Hamiltonian $H_\Psi$  as
\bea
\rho_\Psi \is  \frac{1}{Z_\beta} \spc 
e^{-\beta H_\Psi},   \qquad {Z_\beta} = \Tr \bigl(e^{-\beta H_\Psi}\bigr) 
\eea
The modular Hamiltonian $H_\Psi$ is an element of the algebra ${\cal A}$ and its action on ${\cal A}$ coincides to high accuracy with that of the microscopic CFT Hamiltonian. 
Since the state $\Psi$ is a microcanonical sum \eqref{micropure} centered around a specified energy $E$, we are instructed to fix the inverse temperature $\beta$ via the condition that
\bea
\label{scond}
S(\rho_\Psi) \is \bigl(1- \beta{\partial_\beta}\bigr)  \log Z_\beta =\, S(E),
\eea
with $e^{S(E)}$ the spectral density at energy $E$. This condition will be important in what follows.

The reduced density matrix $\rho_\Phi$ of the Hawking pair state \eqref{hpair} then takes the form
\bea 
\label{rhophi}
\rho_\Phi \is \frac{1}{\zZ_{\beta,\tau\!\!}}\, \spc{\phi}_{\tau} \spc \bar{\phi}_{\tilde\tau}\spc e^{-\beta H_\Psi}
{\phi}_{-\tilde\tau } \spc \bar{\phi}_{-\tau}\ \\[2mm]
\zZ_{\beta,\tau} \is \Tr\Bigl(e^{-\beta H_\Psi} {\phi}_{-\tilde\tau } \spc \bar{\phi}_{-\tau} \spc{\phi}_{\tau} \spc \bar{\phi}_{\tilde\tau}\spc \Bigr)
\eea
This density matrix $\rho_\Phi$ is the unique element of the operator algebra ${\cal A}$ such that $\la \Phi| {\bf a} |\Psi\ra = {\rm Tr}(\rho_\Phi {\bf a})$ for all ${\bf a} \in {\cal A}$. 
We will study the properties of $\rho_\Phi$ in the next section. As a warm-up, it will be useful to first take a closer look at the normalization factor, given by the thermal four point function $\zZ_{\beta,\tau}$. One can think of $\zZ_{\beta,\tau}$ as the thermodynamic partition function of the Hawking pair. 
 
In the large $N$ limit, the bulk QFT is non-interacting and the four point function can be evaluated by means of Wick's theorem. There are two Wick contractions, depicted in equation \eqref{twowick}. 
The boundary-to-boundary propagators in the black hole background coincide with the thermal two-point function~\eqref{thermaltwo}. So, to leading order in large $N$, we find 
\bea
\label{zwicko}
{Z_{\beta,\tau}}/{Z_\beta} \is G_{\beta} (\tau\nspc-\nspc \tilde{\tau}\smpc)^2 + G_{\beta} (\smpc{2\tau}\smpc)^2 
\eea

The holographic Wick rule can be motivated directly from the microscopic CFT perspective. We can write the four-point function $\zZ_{\beta,\tau}$ as follows
\bea
\label{zctrace}
\zZ_{\beta,\tau}\! \is\!  {\rm Tr}\bigl(\nspc \bfC^\dag\nspc \bfC\spc \bfC^\dag\nspc \bfC \bigr), \ \qquad \ \bfC \nspc \equiv \spc  e^{-\frac \beta 4 H_\Phi} \phi_\tau\strut
\eea
In a maximally chaotic CFT, the $\bfC$ operators can be thought as hermitian random matrices. To make this more explicit, let us expand them in an energy eigenbasis
\bea
\label{cmatrix}
\bfC\!\is\!\!\sum_{E,E'}  {C_{E E'}} |E'\rangle\langle E | \quad \qquad
{C_{E E'}}\! = \spc \la E'| \spc e^{-\frac \beta 4 H_\Phi} \phi_\tau |E\ra 
\eea
The matrix elements have random phases subject to the hermiticity condition $C^*_{E E'}\! =\nspc C_{E'E}$. Hence the products $\bfC^\dag\nspc \bfC$
and $\bfC\smpc \bfC^\dag$ that appear inside the trace \eqref{zctrace} are well approximated by diagonal matrices, up to exponentially small off-diagonal terms, see e.g. \cite{vvqec}-\cite{replica1}. 

Equivalently, in the large $N$ regime of interest, we can treat the $\bfC$ as elements of a gaussian random matrix ensemble with diagonal Wick contractions
\bea
\label{cwick}
\wick{\c1{\bfC^\dag} \c1{\bfC}}= G_{\beta} (\smpc\tau-\tilde{\tau}\smpc) e^{-\frac \beta 2 H_\Psi},\ \ &  & \ \
\wick{ \c1{\bfC}\spc \c1{\bfC\,}}{\!}^\dag\! =  G_{\beta} (2\tau) e^{-\frac \beta 2 H_\Psi}.\ \ \
\eea
Applying the Wick rule to the partition function 
\bea
\label{zwickt}
\zZ_{\beta,\tau}\! \is\nspc  \Tr\bigl( \wick{\c1{\bfC}{}^\dag \c1{\bfC}}\spc \wick{\c1{\bfC}{}^\dag \c1{\bfC}}\bigr)  +   \Tr \bigl(\wick{\c2{\bfC}{}^\dag \c1{\bfC}\spc \c1{\bfC}{}^\dag \c2{\bfC}}\bigr)\ \;
\eea
reproduces the bulk formula \eqref{zwicko}.

It is instructive to again write the two terms in \eqref{zwickt} in diagrammatic form, but add labels that indicate the energies of the states that contribute in each channel
\bea
\label{channels}
~~\raisebox{-19.5mm}{\begin{tikzpicture}[scale=.67]);
\draw[thick] (4,0) arc (0:360:2);
\draw[color={rgb:red,10; black,3}, fill={rgb:red,10; black,3}]   (1,-1.7) circle (0.1); 
\draw[color={rgb:red,10; black,3}, fill={rgb:red,10; black,3}]   (3,-1.7) circle (0.1); 
\draw[color={rgb:red,10; black,3}, fill={rgb:red,10; black,3}]   (1,1.7) circle (0.1); 
\draw[color={rgb:red,10; black,3}, fill={rgb:red,10; black,3}]   (3,1.7) circle (0.1); 
\draw[thick, color={rgb:red,10; black,3}] (1,-1.7) arc (90+38:90-38:1.6);
\draw[thick, color={rgb:red,10; black,3}] (1,1.7) arc (270-38:270+38:1.6);
\draw[dashed] (4,0) -- (0,0);
\draw[fill=black] (2,0) circle (0.1);
\draw (-.35,0.05) node {\small $E$};
\draw (2,2.35) node {\small $E'_1$};
\draw (2,-2.35) node {\small $E'_2$};
\draw (4.35,0.05) node {\small $E$};
\end{tikzpicture}}~~~~~~~~~~~~\raisebox{0mm}{\begin{tikzpicture}[scale=.67, rotate=90, baseline={([yshift=0cm]current bounding box.center)}]);
\draw[thick] (4,0) arc (0:360:2);
\draw[color={rgb:blue,10; black,3}, fill={rgb:blue,10; black,3}]   (1,-1.7) circle (0.1); 
\draw[color={rgb:blue,10; black,3}, fill={rgb:blue,10; black,3}]   (3,-1.7) circle (0.1); 
\draw[color={rgb:blue,10; black,3}, fill={rgb:blue,10; black,3}]   (1,1.7) circle (0.1); 
\draw[color={rgb:blue,10; black,3}, fill={rgb:blue,10; black,3}]   (3,1.7) circle (0.1); 
\draw[dashed] (2,2) -- (2,-2);
\draw[fill=black] (2,0) circle (0.1);
\draw[thick, color={rgb:blue,10; black,3}] (1,-1.7) arc (90+38:90-38:1.6);
\draw[thick, color={rgb:blue,10; black,3}] (1,1.7) arc (270-38:270+38:1.6);
\draw (-.35,0) node {\small $E'$};
\draw (2,2.45) node {\small $E_1$};
\draw (2,-2.45) node {\small $E_2$};
\draw (4.35,0) node {\small $E'$};
\end{tikzpicture}}\ 
\eea
We see that segments that are connected through the bulk without crossing any propagators carry the same energy. From the boundary perspective, this reflects the fact that the Wick contractions \eqref{cwick} are diagonal matrices in the energy basis. The other energies with subscript labels are independently summed over. Between the two terms, the one with the shortest boundary-to-boundary geodesic gives the dominant contribution. The CFT mechanism for this is entropic: the channel with the largest number of  intermediate states gives the largest result for the Wick contraction.

\def\hh{{\;\widehat{\!\! h\!}\,}}

\section{{{Entropy of a Hawking pair}}}
\vspace{-1mm}
\setcounter{equation}{5}

We wish to study the entropy difference 
\bea
\label{deltas}
\Delta S \is 
S(\rho_{\Phi\nspc})\nspc -\nspc S(\rho_\Psi)
\eea
between the state $\Psi$ and the state $\Phi$ with the Hawking pair. We will interpret this difference as a measure of the quantum information content of the Hawking pair. Below we will compute $\Delta S$ using three different methods: from thermodynamics, via a semi-classical replica method, and using exact JT gravity results. In the next section we will then relate the result of these computations to the recent insights about the role of type II${}_\infty$ operator algebras in semi-classical holography \cite{LL1}-\cite{wittenetal}. 
\setcounter{equation}{32}

\subsection{{{$\Delta S$ from Thermodynamics}}}
\vspace{-.5mm}

\def\sss{{\beta(1-n)}\spc}

As figure \eqref{channels} indicates, we can assign two thermodynamic energies  $E$ and $E'$ to the Hawking pair state~$\Phi$. Their dependence on  $\beta$ and $\tau$ follows by taking the variation with respect to $\beta$, while keeping either $x=2{\tau}$ or $\tilde{x}=\tau\!-\!\tilde\tau$ held fixed. A straightforward calculation shows that
\bea
\biggl( \frac{\raisebox{-1pt}{$\partial F_{\beta,\tau}\!$}}{\raisebox{-.5pt}{$\partial \beta$}} 
\nspc \biggr)_{\!\tilde{x}}\! \is E\qquad \ \
\biggl( \frac{\raisebox{-1pt}{$\partial F_{\beta,\tau}\!$}}{\raisebox{-.5pt}{$\partial \beta$}} 
\nspc \biggr)_{\! x}\! = \, E', \ 
\eea
with  $F_{\beta,\tau} = -\log \zZ_{\beta,\nspc\tau}$. Plugging in \eqref{twocases}~gives
\bea
\label{edef}
\ \ E
\is \! E_\beta\nspc -\nspc  \partial_\beta 
\nspc \log G_{\beta}(\tau\!-\nspc \tilde{\tau})^2 \qquad\    \; I \\[2.5mm]
\ \ E'\!
\is\spc E_\beta  -  \partial_\beta
\nspc  \log G_{\beta}(\smpc 2\tau \smpc)^2  \qquad \  \ \; I\nspc I
\label{epdef}
\eea
with $E_\beta = -\partial_\beta\log Z_\beta$. Here the partial $\beta$-derivative in \eqref{edef} is defined with $\tilde{x}$ = constant, while in \eqref{epdef} it is defined with $x$ = constant.

$E$ in \eqref{edef} is the energy of the intermediate state indicated in the left diagram in \eqref{channels}. It represents the energy of the initial black hole state $\Psi$.  In our micro-canonical setting,  we should read \eqref{edef} as an equation that determines how $\beta$ should vary with $\tau$ so that $E$ remains constant.  $E'$ in \eqref{epdef} is the energy of the intermediate black hole state between $\phi_\tau$ and $\phi_{\tilde\tau}$ in the right diagram in \eqref{channels}. It represents the energy of the black hole after the Hawking particle has come out.  This energy $E'$ varies with $\tau$, while $E$ is kept fixed. 
We will view $E$ and $E'$ in \eqref{edef}-\eqref{epdef} as defined over the full range of $\tau$, even though they are computed from different Wick contraction components of the four point function ${Z}_{\beta,\tau}$.

To obtain the entropy of the Hawking pair state $\Phi$ as a function of the particle location $\tau$ we apply the first law
\bea
\Delta S \is \beta \Delta E\spc = \spc \beta \bigl(E' \nspc - E)
\eea
Here $\Delta E$ is the change in energy of the black hole state due to the emission of the Hawking particle, given that it was detected at location~$\tau$. $\Delta S$ is the amount of entropy that has escaped from the black hole along with the Hawking particle. Plugging in equations \eqref{edef} and \eqref{epdef} gives the result \eqref{deltasfin} announced in the introduction.

\begin{figure}[t]
\begin{center}
\begin{tikzpicture}[xscale =4,yscale = 1.6,domain=0.01:pi/2-0.01]
  \draw[->] (0, 0) -- (1.7, 0) node[right] {$r$~~~~${}$};
  \draw[->] (0, -1) -- (0, 1.05) node[left] {${E'}$};
  \draw[dashed] (.78, -.75) node[below] {\small $r\!=\! r_c$} -- (.78,1);
  \draw[dashed] (1.57, -1)-- (1.57,1)node[right] {${E}$};     
  \draw[thick,cyan]   plot (\x,{(\x*cot(\x  r) - (pi/2-\x)*tan(\x r)) }) ;
  \matrix [draw,lightgray,below right] at (current bounding box.north east) { \\
   \node [color={rgb:red,5; black,1}]  {\small $-\!\!\!- \; E-E_\beta$}; \\
   \node [color={rgb:blue,10; black,1}]  {\small $-\!\!\!- \; E'-E_\beta$}; \\
  \node [cyan] {\small $-\!\!\!-\;  E' - E$}; \\[1mm]
  \node [color={rgb:black,5; black,5}] {$\mbox{\small{\bf{-\;\!-\;\!-} }}$\, $\Delta S/\beta$};\\
};
\draw (1.12,1) node{$I\nspc I$};
\draw (.44,1) node{$I$};
    \draw[thick,color=blue]   plot (\x,{-(pi/2-\x)*tan(\x r) +1})  ;
    \draw[thick,color=red]   plot (\x,{-\x*cot(\x  r) +1 }) ;
    \draw[very thick,color={rgb:black,5; black,5}, dashed, domain=pi/4+0.01:pi/2-.01]   plot (\x,{(\x*cot(\x  r) - (pi/2-\x)*tan(\x r)-.03) }) ;
    \draw[very thick,color={rgb:black,5; black,5}, dashed, domain=0:pi/4]   plot (\x,{-.03}) ;
    \draw[very thick,color={rgb:black,5; black,5}, dashed, domain=pi/4+0.02:pi/2-.01]   plot (\x,{(\x*cot(\x  r) - (pi/2-\x)*tan(\x r)-.03) }) ;
    \draw[very thick,color={rgb:black,5; black,5}, dashed, domain=0.01:pi/4]   plot (\x,{-.03}) ;
    \draw[very thick,color={rgb:black,5; black,5}, dashed, domain=pi/4+0.01:pi/2-.01]   plot (\x,{(\x*cot(\x  r) - (pi/2-\x)*tan(\x r)-.025) }) ;
    \draw[very thick,color={rgb:black,5; black,5}, dashed, domain=0:pi/4]   plot (\x,{-.025}) ;
    \draw[very thick,color={rgb:black,5; black,5}, dashed, domain=pi/4+0.02:pi/2-.01]   plot (\x,{(\x*cot(\x  r) - (pi/2-\x)*tan(\x r)-.025) }) ;
    \draw[very thick,color={rgb:black,5; black,5}, dashed, domain=0.01:pi/4]   plot (\x,{-.025}) ;
\end{tikzpicture}
\vspace{-3mm}
\end{center}
\caption{The red and blue graph depict the energy $E$ and $E'$ as a function of $\tau$ running from 0 to $\beta/4$, with $\beta$ kept constant. The entropy difference $\Delta S$ vanishes in the region where $E'>E$ and equals $\beta \Delta E$ in the region where $E'<E$. }
\vspace{-1mm}
\end{figure}
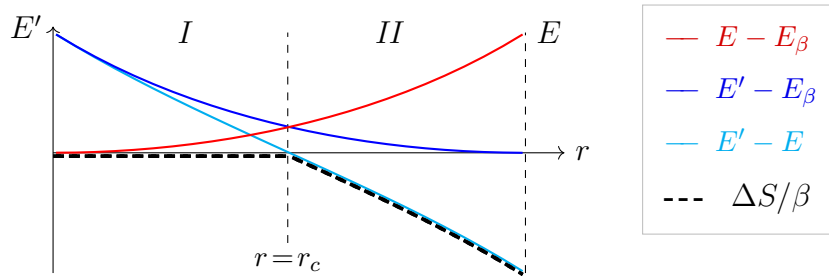

To make our result a bit more explicit, let us consider the case when the boundary theory is a 1D or chiral 2D CFT. The thermal two-point function of a primary operator with scale dimension $h$ then takes the~simple form
\bea
\label{greeno}
 G_{\beta}( 2\tau) \is  \left(\frac{{2\pi}/{\beta}}{\sin\bigl(2\pi \tau/\beta\bigr)}\right)^{2h}
\eea
Plugging this explicit form into equations \eqref{edef}-\eqref{epdef} gives
\bea\qquad E
\, = & &\!\!\!\!\!\!\!\!\! \nspc E_\beta \nspc -\nspc \frac{4h}{\beta} \biggl(\nspc 1-\frac{\pi (\tau\!-\! \tilde\tau)}{\beta} \nspc \cot\Bigr(\frac{\pi\smpc(\tau\!-\!\tilde\tau)}{\beta}\Bigr)\nspc\biggr),\quad 
\label{erel}
\\[2mm]
\label{eprel}
 E'
\! \is\! E_\beta  -  \frac{4h}{\beta}\biggl(\nspc 1- \frac{2\pi \tau}{\beta} \nspc \cot \Bigr(\frac{2\pi \tau}{\beta}\Bigr)\nspc\biggr).
\eea

We plot both functions in figure 2. The red and blue graphs depict $E$ and $E'$ as functions of $\tau$ with $\beta$ held constant.  In the near horizon region $I$, $E'$ exceeds~$E$, while in the far away region $I\nspc I$, $E'$ is smaller than~$E$. The entropy difference $\Delta S$ vanishes in region $I$ and equals $\beta \Delta E$ in the region $I\nspc I$, where both quantities are negative. This is the regime where the entropy extraction takes place. The explicit formula for $\Delta S$ in region $I\nspc I$ reads
\bea
\label{deltasfint}
\Delta S \nspc \is \beta (E'-E)\, = \, {4 h} \biggl(\frac{4 \pi \tau}{\beta} {\rm cosec} \Bigl(\frac{4 \pi \tau}{\beta}\nsmpc \Bigr)\! -\!
 \frac{\pi}2 \tan \Bigr(\nsmpc \frac{2 \pi \tau}{\beta}\nsmpc \Bigr)
  \biggr)
 \eea
 We see that $\Delta S$ is proportional to the scale dimension $h$ of the primary operator $\phi_\tau$. This matches the physical expectation that the entropy extracted from the black hole is proportional to the mass of the Hawking particle.\footnote{One can make this comparison more precise in the 2D case. In a BTZ background \eqref{btz}, a $\phi$ particle at rest at radial location $r=\tau +\beta/4$ would carry energy $\Delta E_{\rm particle} = 2h \frac{2\pi}{\beta} \cot \frac{2\pi \tau}{\beta}$. Here the $\tau$ dependence comes from the Tolman factor (see section 2.2, equation \eqref{btz} and the upcoming paper \cite{toappearsoon} for more discussion). While $\Delta S/\beta$ and $\Delta E_{\rm particle}$ scale the same way with $h$, our calculation shows that 
 \bea
 \Delta S/\beta = \Delta E < \Delta E_{\rm particle} 
 \eea
 We see that only a fraction of the rest energy of a detected Hawking particle in fact gets extracted from the black hole. Evaporation of a BTZ black hole requires an external source of energy in the form of a detector. We will return to this point in the concluding section.}

\subsection{{$\Delta S$ from Replica Method}}
\vspace{-.5mm}

The replica method for computing the von Neumann entropy of $\rho_\Phi$ exploits that it arises as the limit
\bea\label{renyie}
S(\rho_{\Phi}) \is \lim_{n\to 1} S_n(\rho_{\Phi}) 
\eea
of the $n$-th R\'enyi entropy. Following \cite{Dong}, we will adopt the following slightly non-standard definition  of the R\'enyi entropy 
\bea
\label{renyin}
S_n(\rho_{\Phi})\! \is\!
\bigl(1\! - \! n \partial_n 
\bigr) 
\log\Tr\bigl(\rho_\Phi^{n}\bigr).
\eea
This definition coincides with the standard one in the $n\to 1 $ limit and turns out to be better attuned to holographic replica calculations.

Inserting the explicit form \eqref{rhophi} of $\rho_\Phi$, we find that
\bea
\Tr\bigl( \rho_\Phi^{n}\bigr) \! \is\!  \frac 1 {\zZ^{n}_{\beta,\tau\!\!}} \, \Tr\Bigl(e^{-n\beta H_\Psi}\!\!
\prod_{k=0}^{2n-1} \! \spc {\phi}{}_{\nspc{\raisebox{-1.5pt}{\scriptsize $\tau  \! + \! k \nspc  \frac \beta 2$} }}\spc
\bar{\phi}_{{\raisebox{-1.25pt}{\scriptsize $\tilde\tau\! +\! k \frac \beta 2$}}}\Bigr)
\eea
where the product is time ordered. 
Applying the definition \eqref{zctrace} of the $C$ matrix, we can rewrite this as
\bea
{\rm Tr}\bigl(\rho_\Phi^n\bigr)\! \is \!  \frac{1}{\! {Z}_{\beta,\tau}^n\!}\; {\rm Tr}\Bigl(\raisebox{-3.5mm}{\;${\underbrace{\!\! {\bfC}^\dag\!{\bfC} {\bfC}^\dag{\bfC}\; \ldots\; {\bfC}^\dag \!{\bfC} {\bfC}^\dag\!{\bfC}\strut\!\!}}\atop{\mbox{$n$ times}}$}\;\Bigr)\
\eea
As before, we can evaluate both expressions by applying Wick's theorem, either by using the bulk free field theory or the contraction rule \eqref{cwick} of the $\bfC$ matrices. This produces a sum of many terms given by all the possible Wick contractions. However, it is clear that, depending on which of the two regimes we are in, the dominant contribution comes from one of the following two types of diagrams with only nearest neighbor contractions (here drawn for the 2$^{\rm nd}$ R\'enyi entropy, $n=2$)
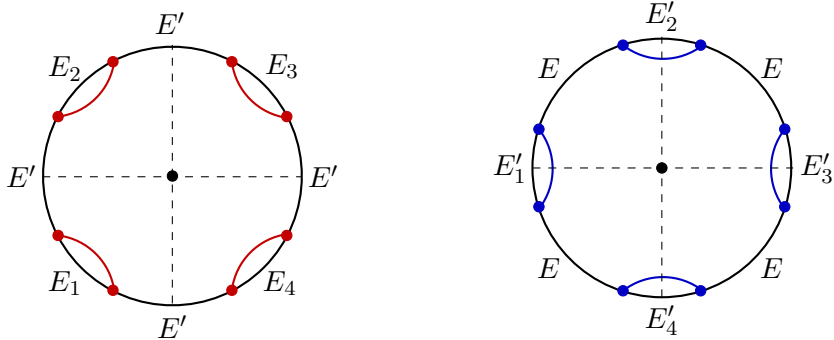
\begin{figure}[hbtp]
\begin{center}
\raisebox{.15mm}{\begin{tikzpicture}[scale=.86, rotate=45]
\draw[thick, color={rgb:red,10; black,3}] (-5.9,0.6) arc (38:-38:1);
\draw[thick, color={rgb:red,10; black,3}] (-5.9+3.8,0.6) arc (180-38:180+38:1);
\draw[thick, color={rgb:red,10; black,3}] (-4-0.6,1.9) arc (270-38:270+38:1);
\draw[thick, color={rgb:red,10; black,3}] (-4-0.6,-1.9) arc (90+38:90-38:1);
\draw[dashed,rotate=-45] (-5.84+1,-2.84) -- (-1.84+1,-2.84);
\draw[dashed, rotate=-45] (-4+1.17,-2-2.8) -- (-4+1.17,2-2.8);
\draw[thick] (-2,0) arc (360:0:2);
\draw[fill=black] (-4,0) circle (0.08);
\draw[color={rgb:red,10; black,3}, fill={rgb:red,10; black,3}] (-5.9,0.6) circle (0.08);
\draw[color={rgb:red,10; black,3}, fill={rgb:red,10; black,3}] (-5.9+3.8,0.6) circle (0.08);
\draw[color={rgb:red,10; black,3}, fill={rgb:red,10; black,3}] (-5.9,-0.6) circle (0.08);
\draw[color={rgb:red,10; black,3}, fill={rgb:red,10; black,3}] (-5.9+3.8,-0.6) circle (0.08);
\draw[color={rgb:red,10; black,3}, fill={rgb:red,10; black,3}] (-4-0.6,1.9) circle (0.08);
\draw[color={rgb:red,10; black,3}, fill={rgb:red,10; black,3}] (-4+0.6,1.9) circle (0.08);
\draw[color={rgb:red,10; black,3}, fill={rgb:red,10; black,3}] (-4-0.6,-1.9) circle (0.08);
\draw[color={rgb:red,10; black,3}, fill={rgb:red,10; black,3}] (-4+0.6,-1.9) circle (0.08);
\draw (-6.35,0) node {\small $E_1$};
\draw (-4,2.35) node {\small $E_2$};
\draw (-4,-2.35) node {\small $E_4$};
\draw (-1.6,0) node {\small $E_3$};
\draw (-5.65,1.65) node {\small $E'$};
\draw (-5.65,-1.65) node {\small $E'$};
\draw (-2.39,1.66) node {\small $E'$\strut};
\draw (-2.35,-1.65) node {\small $E'$};
\end{tikzpicture}}
\hspace{1.5cm}
\begin{tikzpicture}[scale=.86]
\draw[thick, color={rgb:blue,10; black,3}] (-5.9,0.6) arc (38:-38:1);
\draw[thick, color={rgb:blue,10; black,3}] (-5.9+3.8,0.6) arc (180-38:180+38:1);
\draw[thick, color={rgb:blue,10; black,3}] (-4-0.6,1.9) arc (270-38:270+38:1);
\draw[thick, color={rgb:blue,10; black,3}] (-4-0.6,-1.9) arc (90+38:90-38:1);
\draw[dashed] (-6,0) -- (-1.97,0);
\draw[dashed] (-4,-2) -- (-4,2.03);
\draw[thick] (-2,0) arc (360:0:2);
\draw[fill=black] (-4,0) circle (0.08);
\draw[color={rgb:blue,10; black,3}, fill={rgb:blue,10; black,3}] (-5.9,0.6) circle (0.08);
\draw[color={rgb:blue,10; black,3}, fill={rgb:blue,10; black,3}] (-5.9+3.8,0.6) circle (0.08);
\draw[color={rgb:blue,10; black,3}, fill={rgb:blue,10; black,3}] (-5.9,-0.6) circle (0.08);
\draw[color={rgb:blue,10; black,3}, fill={rgb:blue,10; black,3}] (-5.9+3.8,-0.6) circle (0.08);
\draw[color={rgb:blue,10; black,3}, fill={rgb:blue,10; black,3}] (-4-0.6,1.9) circle (0.08);
\draw[color={rgb:blue,10; black,3}, fill={rgb:blue,10; black,3}] (-4+0.6,1.9) circle (0.08);
\draw[color={rgb:blue,10; black,3}, fill={rgb:blue,10; black,3}] (-4-0.6,-1.9) circle (0.08);
\draw[color={rgb:blue,10; black,3}, fill={rgb:blue,10; black,3}] (-4+0.6,-1.9) circle (0.08);
\draw (-6.35,0) node {\small $E'_1$};
\draw (-4,2.35) node {\small $E'_2$};
\draw (-4,-2.35) node {\small $E'_4$\strut};
\draw (-1.6,0) node {\small $E'_3$};
\draw (-5.75,1.55) node {\small $E$};
\draw (-5.75,-1.55) node {\small $E$};
\draw (-2.3,1.55) node {\small $E$};
\draw (-2.3,-1.55) node {\small $E$};
\end{tikzpicture}
\vspace{-5mm}
\end{center}
\end{figure}

The red diagram dominates in region $I$ and the blue diagram dominates in region $I\nspc I$. Any other diagrams involve boundary-to-boundary propagators associated with longer geodesics and are therefore exponentially suppressed as soon as one moves away from the transition point $\tau = \tau_c$ between the two regions.  We thus find that
\bea
\label{rhone}
\Tr\bigl(\rho_\Phi^{n}\bigr)
\is  \frac{Z_{n\beta}}{Z_\beta^n} \biggl(\frac{G_{n\beta}(y)}{G_{\beta}(y)}\biggr)^{\! 2n}
\ ; \ \ \ y = \min(2\tau,\tau\!-\! \tilde\tau)
\eea
Inserting this expression into equation \eqref{renyin} gives that the entropy of the Hawking pair state $\Phi$ in the near horizon region $I$ equals
 \bea
\ \ S(\rho_\Phi)_I\nspc \is\nspc  \bigl(1\nspc-\nspc n  
\partial_n \bigr)  \nspc \log \Bigl(Z_{n\beta} \spc G_{n\beta} (\tau\!-\! \tilde\tau)^{2n}\Bigr){}_{|_{\strut}n=1} \eea
Introducing  the thermal entropy at energy $E_\beta$ via $S(E_\beta) = (1 - \beta\partial_\beta) \log Z_\beta$ we compute
\bea
\label{first}
S(\rho_\Phi)_I
\is  S(E_\beta) \!-\nspc \beta 
\partial_\beta\nspc \log G_{\beta} (\tau\!-\! \tilde\tau)^2\ \ \nn \\[3mm]
\is  S(E_\beta) +\beta(E-E_\beta)\\[3mm]
\is S(E)\, = \, S(\rho_\Psi)\nn
\eea
In the first line, the partial derivative with respect to $\beta$  is defined with $\tau\nspc - \nspc\tilde\tau$ kept constant. To go to the second line, we used the relation \eqref{edef} between $E$ and $E_\beta$ and in the final steps we used  that $\beta = {dS}/{dE}$ and the condition~\eqref{scond}.  We again find that, in the leading semi-classical limit, the entropy difference between $\Phi$ and $\Psi$  vanishes in region~$I$. 

The entropy difference $\Delta S = S(\rho_\Phi)_{I\nspc I} - S(\rho_\Psi)$ in region $I\nspc I$
is non-vanishing. 
Using equations \eqref{renyin} and \eqref{rhone}, we obtain the result \eqref{deltasfin} announced in the Introduction.
Plugging in the explicit form \eqref{greeno} of the 1D thermal two point functions also reproduces the 1D result \eqref{deltasfint}.

\bigskip

\subsection{{$\Delta S$ from JT gravity}}

\vspace{-1mm}

The coarse-grained mechanism of entropy extraction is simply that gravitational backreaction due to presence of the Hawking pair reduces the black hole mass and horizon area. The Bekenstein-Hawking formula then stipulates that entropy of the black hole is reduced and transferred to the exterior. Our computation quantifies how the horizon area varies as a function of the location of the Hawking pair. In Appendix A, we explicitly verify this gravitational interpretation in the context of JT gravity.

\pagebreak

\section{{Hawking pair entropy from type II$_\infty$}}

\vspace{-2mm}

The replica calculation of the Hawking pair entropy relies on two basic principles: the free field Wick rule and the postulate that there exist density matrices  $\rho_\Phi$ and $\rho_\Psi$ that describe the black hole with or without the Hawking pair. These two assumptions were enough to conclude that the pair can extract entropy from the black hole. This stands in contrast with the standard QFT in a fixed curved-spacetime treatment of Hawking evaporation. This raises the question: where exactly do the two calculations differ?  
Here we will address this question using the recent insights into the role of von Neumann algebras in holography. 

As pointed out in \cite{wittenx}, the semi-classical large $N$ limit of a holographic CFT  defines a deformation from pure QFT in a fixed background, turning it from a type III von Neumann algebra into a type II$_\infty$ algebra. In the type III setting, quantum information can get lost due to the fact that the fluctuations in the candidate modular Hamiltonian of the QFT diverge at the horizon. In \cite{wittenx}, the algebra of single-trace operators in a large $N$ CFT is identified with the so-called crossed product of the undeformed type III operator algebra ${\cal A}_{\rm qft}$ with the one parameter modular automorphism group generated by the modular Hamiltonian $h_\psi$. We briefly summarize the construction of \cite{wittenx}-\cite{wittenetal2}.

\def\alphaE{\alpha}

The modular time flow generated by the modular Hamiltonian $h_\Psi$ annihilates the state $\Psi$. This tells us that the associate timeflow is a Killing symmetry of the bulk spacetime. The time flow of the extended black hole spacetime runs forward in the exterior region R but backwards in the interior region L. So, schematically, we can write $h_\Psi$ as the difference between the right and the left Hamiltonian $h_\Psi = h_R - h_L$. 
Let $q$ be the generator of the time evolution in $t_L+t_R$, given by the sum of the interior and exterior Hamiltonian $q = h_L+h_R.$
 This time flow can not be measured by an outside observer and $q$ is therefore not an element of the QFT operator algebra ${\cal A}_{\rm qft}$.  The crossed product is defined as the extended algebra $\widehat{\cal A}$ generated by all QFT operators ${\bf a} \in {\cal A}_{\rm qft}$ along with the generator  
\bea
\hhR\! &\!\equiv \! &\! (h_\Psi  +q)/2,\ \
\eea
with $h_\Psi$ the modular Hamiltonian of the state $\Psi$ defined in equation \eqref{modularo}. This algebra acts on the Hilbert space ${\cal H}_{\rm cr} = {\cal H}_{\rm qft} \otimes L^2(\mathbb{R})$. As explained in \cite{wittenetal}, this crossed product defines a type II$_\infty$ algebra $\widehat{\cal A}$. In particular, it admits the notion of a density matrix. 

The authors of \cite{wittenetal,wittenetal2} introduce a simple product state of the form $\hhat{\Psi} = \Psi \otimes f(q) \in {\cal H}_{\rm cr}$, with $f(q) = \epsilon^{1/2} g(\epsilon q)$ with $g(q)$ some smooth $L^2$ function. The function $f(q)$ can be thought of as the pure state wave function of an auxiliary clock, with time resolution of order $\epsilon$, that keeps track of the time flow generated by the extra operator $q$.\footnote{More generally, one could consider entangled states $\hhat\Psi = \sum_i \Psi_i \otimes f_i(q)$ between the CFT and the clock. After tracing over the clock, this produces a state of the thermal mixed double form.} The density matrix $\rho_{\hhat\Psi} \in \widehat{\cal A}$ corresponding to $\hhat\Psi$ takes the following form
\bea
\label{rpsiansatz}
\ \rho_{\hhat\Psi} \is  |{f}(\hhR)|^2 e^{-\beta\spc\hhR}.
\eea 
In the small $\epsilon$ limit, the function $f(q)$ reduces to a constant and the state \eqref{rphiansatz} approaches the thermal density matrix  with inverse temperature $\beta$ associated with $\hhR$.

To write the density matrix of the Hawking pair state $\hhat\Phi$ in this formalism, we introduce the relative modular operator $\Delta_{\Phi|\Psi}$ and hamiltonian $h_{\Phi|\Psi}$ via
\bea
\Delta_{\Phi|\Psi} \is S^\dag_{\Phi|\Psi}S_{\Phi|\Psi} = \, e^{-\beta h_{\Phi|\Psi}}
\eea
where $S_{\Phi|\Psi}$  is the relative Tomita operator, defined as the anti-linear map with the property that $S_{\Phi|\Psi} {\bf a} \Psi = {\bf a}^\dag \Phi$ for all ${\bf a} \in {\cal A}_{\rm qft}$. The Connes cocycle flow defined as
\bea
\label{connes}
 u_{\Phi|\Psi}(s)  \is \Delta^{is}_{\Phi|\Psi} \Delta^{-is}_\Psi \, =\,  \Delta^{is}_\Phi \Delta^{-is}_{\Psi|\Phi} 
\eea
is valued in the QFT algebra ${\cal A}_{\rm qft}$ if $s$ is real. The above equality implies the following relation 
among the modular Hamiltonians
\bea
\label{hrel}
\beta (h_\Psi-h_{\Phi|\Psi}) \is -\beta (h_\Phi - h_{\Psi|\Phi})
\eea
Each side of \eqref{hrel} defines an operator that is affiliated with the QFT operator algebra ${\cal A}_{\rm qft}$.

Applying the construction of \cite{wittenetal}, we now write the density matrix $\rho_{\hhat\Phi}\in \widehat{\cal A} $ associated the Hawking pair state $\hhat\Phi$ as 
\bea
\label{rphiansatz}
  \!\! \rho_{\hhat\Phi} \is  |{f}(\hhR)|^2 e^{-\beta (h_R - h_\Psi + h_{\Phi|\Psi})}
\eea
This formula reduces to \eqref{rphiansatz} in case $\hhat\Phi$ equals $\hhat\Psi$. Since $\hhR - (h_\Psi - h_{\Phi|\Psi})$ is a sum of two terms affiliated with ${\widehat{\cal A}}$,
 the expression \eqref{rphiansatz} has the right properties to be a candidate density matrix of the state $\hhat{\Phi}$ for the algebra $\widehat{\cal A}$.  In \cite{wittenetal} it is further shown that $\Tr(\rho_{\hhat\Phi} {\bf a}) = \la \hhat\Phi| {\bf a} |\hhat\Phi\ra$ for all ${\bf a} \in \widehat{\cal A}$, confirming the proposed identification.

The operator formalism is well suited for identifying the physical difference between the holographic setting and QFT in a fixed curved spacetime.  It will therefore be instructive to consider the entropy difference $\Delta S = - \Tr\rho_{\hhat\Phi}  \log \rho_{\hhat\Phi}  + \Tr \rho_{\hhat\Psi} \log \rho_{\hhat\Psi}$
between the states $\hhat\Phi$ with $\hhat\Psi$ with and without the Hawking pair in this set up.  Plugging in the expressions \eqref{rphiansatz} and \eqref{rpsiansatz} we find that\footnote{
Here we assumed that the origin of the $q$ variable is shifted so that $\la \hhat\Psi| q |\hhat\Psi\ra = 0$.} $
\Delta S= \la \Phi | h_{\Phi|\Psi} |\Phi\ra - \la \hhat\Phi| \beta q |\hhat\Phi\ra$.
Using the relation \eqref{hrel} and that $\la \Phi|h_\Phi|\Phi\ra=0$, we can rewrite this as the difference 
\bea 
\Delta S 
\is \nspc S_{\rm grav}(\Phi,\Psi)\nspc -\nspc S_{\rm rel}(\Phi|\!\smpc |\Psi)_{\strut}
\eea
between the gravitational entropy contribution and the relative entropy
\bea  
\label{sgrav}
 S_{\rm grav}(\Phi,\Psi)\nspc \is\smpc 
 \la \smpc \hhat\Phi\smpc |\smpc \beta \hhR  |\smpc \hhat\Phi\smpc \ra\ 
\\[2.5mm]
\label{srel}
S_{\rm rel}(\Phi\smpc |\!\smpc|\smpc\Psi) \is 
\la \smpc \Phi\smpc |\smpc  h_{\Psi|\Phi}  |\smpc \Phi\smpc \ra\ 
\eea
We will now describe how to compute both terms via a replica method.

\subsection{Computation of $S_{\rm grav}$}

\vspace{-1mm}

The Hawking pair state $\hhat\Phi$ is obtained by acting with the local operator $\phi_\tau$ and its mirror  $\tilde\phi_\tau$ on the black hole state $\hhat\Psi$. The local operator $\phi_\tau$ that creates the exterior Hawking particle is, to leading order, part of the exterior operator algebra $\widehat{\cal A}$. In the semi-classical holographic setting, we need to specify a gravitational dressing that anchors the operator to the exterior boundary.  Equivalently, we demand that exterior operators commute with the hamiltonian $\hhL = (q-h_\psi)/2$ that generates the interior time flow  
\bea
\label{phys}
 [\, \hhL, \phi_\tau]
\nspc\is\nspc 0.
\eea
Solutions to this condition take the form $\phi_\tau = e^{ip\smpc h_\Psi} \phi^{\rm qft}_\tau  e^{-ip \smpc h_\Psi}$ with $\phi^{\rm qft}_\tau \in {\cal A}_{\rm qft}$ and $p$ the canonical conjugate time dual to the energy variable $q$. The mirror operator $\bar\phi_{\tilde \tau} = J_\Psi \phi_\tau J_\Psi$, on the other hand, acts on the black hole interior and would therefore be expected to commute with the generator $\hhR = (q+h_\Psi)/2$ of the exterior time flow 
\bea
\label{physt}
 [\, \hhR, \bar\phi_{\tilde\tau}\spc]
 \nspc\is\nspc 0.
\eea
This condition is solved by $ \bar\phi_{\tilde\tau} = e^{-ip\smpc h_\Psi} \bar\phi^{\rm qft}_{\tilde\tau} e^{ip \smpc h_\Psi}$.
This relation indeed follows from the fact that the mirror operation $J_\Psi$ is anti-linear and reverses the direction of the modular flow generated by $h_\Psi$.

To find the gravitational contribution $\Delta S_{\rm grav}$ to the entropy difference, we first evaluate
\bea
\label{exps}
\la \hhat\Phi| e^{\sss \spc\!\hhR} |\hhat\Phi\nspc\ra  \is \frac 1 {{Z}_{\beta,\tau}\!}\, \la \hhat\Psi|  
{\phi}{}_{-\tilde\tau }\spc
{\bar\phi}_{{-\tau\nspc}} \spc e^{\sss\spc\!\hhR}\nspc
{\phi}_{\tau}\spc
{{\bar\phi}_{\tilde\tau }}|\hhat\Psi\ra 
\\[2.5mm]
\is \frac 1 {{Z}_{\beta,\tau}\!}\, \la \hhat\Psi| e^{(1-n)\beta\hhR} \spc {\bar\phi}_{{-\tau}} \,
{\phi}_{\spc\tau}\,
{\bar{\phi}_{\spc\tilde\tau }} \,  {\phi}{}_{\beta-\tilde\tau\nspc}|\hhat\Psi\ra. 
\label{second}
\eea
In the first line we inserted the definition of the Hawking pair state $\widehat\Phi$. In going to the second line, we used the KMS condition \eqref{kms} and physical  condition \eqref{phys}. 

The expressions  \eqref{exps}  and \eqref{second} are valid over the whole range of $\tau$. However, each is naturally adapted to the application of one of the two  Wick contraction rules
\bea 
\qquad \ \ \wick{\c1{\,{\phi}}_{\tau}\, \c1{{\bar\phi}_{\spc\tilde\tau }}}\is \wick{\c1{{\phi}}{}_{-\tilde\tau }\, \c1{{\bar\phi}}_{-\tau}}\smpc = \smpc G_{n\beta}(\tau\!-\!\tilde{\tau}), 
\nn\\[-2mm]\\[-2mm]\nn
\qquad \ \ \wick{\c1{
{\bar\phi}}{}_{-\tau} \spc\!\c1{\,
{\phi}}{}_{\tau}}\is \wick{\c1{{\bar\phi}_{\spc \tilde\tau }} \; \c1{{\phi}\;\,}{\!\!}_{\beta-\tilde\tau\nspc}} \, = \, G_{n\beta}(\smpc
2\tau \smpc). 
\eea
Both quantities \eqref{exps}  and \eqref{second}  looks like a thermal fourpoint point function at inverse temperature $n\beta$, or equivalently, evaluated in the euclidean black hole geometry with timelike periodicity $n\beta$. This spacetime looks identical to the replica geometry used in section 4.2, except that in this case the operator insertions are not replicated. The only quantity that is replica sensitive is the inverse temperature.

\begin{figure}[hbtp]
\begin{center}
\begin{tikzpicture}[xscale =4.3,yscale = 2.65,domain=0.01:pi/2-0.01]
  \draw[->] (0, 0) -- (1.7, 0) node[right] {$r$~~~~${}$};
  \draw[->] (0, -.2) -- (0, 1.05) node[left] {${E'}$};
  \draw[dashed] ( .78,-.15) -- (.78,.78) node[above] {\small $r\!=\! r_c$};
  \draw[dashed] (1.57, -.15)-- (1.57,1)node[right] {${E}$}; 
    \draw[very thick,color={rgb:black,5; black,5}, dashed, domain=0.01:pi/4-.01]   plot (\x, {1-(\x*cot(\x  r)-.02});
    \draw[very thick,color={rgb:black,5; black,5}, dashed, domain= pi/4:pi/2]   plot (\x,{1-(pi/2-\x)*tan(\x r)+.02) }) ;
  \matrix [draw,lightgray,below right] at (current bounding box.north east) { \\
   \node [color={rgb:red,5; black,1}]  {\small $-\!\!\!- \;\; E$}; \\
   \node [color={rgb:blue,10; black,1}]  {\small $-\!\!\!- \;\; E'$}; \\
  \node [color={rgb:black,5; black,5}] {\small -\;\!-\;\!- $\Delta S_{\rm grav}/\beta$};\\
};
\draw (1.12,1) node{$I\nspc I$};
\draw (.44,1) node{$I$};
    \draw[thick,color=blue]   plot (\x,{-(pi/2-\x)*tan(\x r) +1})  ;
    \draw[thick,color=red]   plot (\x,{-\x*cot(\x  r) +1 }) ;
\end{tikzpicture}
\vspace{-5mm}
\end{center}
\caption{The entropy difference $\Delta S_{\rm grav}$ at fixed temperature rises in region $I$ where $E'>E$ and decreases to zero in the region where $E'<E$. }
\vspace{-3mm}
\end{figure}
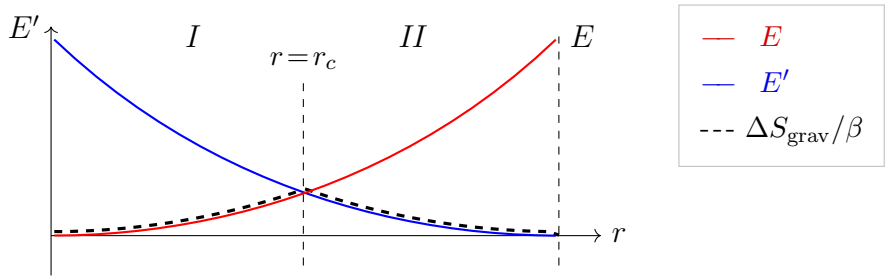

Evaluating \eqref{exps} and \eqref{second} via the respective Wick rule and differentiating the result for \eqref{exps} with respect to $n$, we obtain that
\bea
\label{deltascan}
\qquad \quad\; S_{\rm grav}(\Phi,\Psi)\! \is\!   - {n \frac{\partial\ }{\partial n} \nspc \log\bigl( {G_{n\beta}(\spc y\spc)}^{2}\bigr)_{\!\bigl|n=1}} = -\spc {4h} \biggl(\nspc 1- \frac{\pi y}{\beta} \nspc \cot \Bigr(\frac{\pi y}{\beta}\Bigr)\nspc\biggr)
\eea
with $y = \min(2\tau,\tau\nspc - \tilde\tau)$. Here in the second line we inserted the explicit form \eqref{greeno} of the one-dimensional thermal two-point function. 
This gravitational contribution matches with our result for the entropy of the Hawking pair in the canonical ensemble with fixed $\beta$ computed via the other three methods. As explained in section 3.1, if instead  of a canonical ensemble we use, as dictated by the choice \eqref{micropure} of our state $\Psi$,  the micro-canonical prescription in which $\beta$ is adjusted so that $E$ remains constant, we reproduce our main result \eqref{deltasfin}. Note that the gravitational contribution \eqref{deltascan}  to the entropy difference computed {in the canonical ensemble} is always positive: the presence of the Hawking pair adds extra entropy. This behavior stands in contrast to the entropy difference in the micro-canonical ensemble, which is always negative. As a function of $\tau$, the canonical entropy difference follows a Page-like curve, following the lower segments of the red and blue graph shown in figure 3.  $S_{\rm grav}$ first rises until the cricital value $\tau=\tau_c$ and then decreases down back to zero. We will comment more on this Page-like behavior in the concluding section.

\subsection{Computation of $S_{\rm rel}$}

\vspace{-1mm}

The Araki formula \eqref{srel} for the relative entropy between $\Phi$ and $\Psi$
 holds for general von Neumann algebras, including type III. Indeed, equation \eqref{srel} does not involve the extra generator $q$ and only refers to the QFT operator algebra. This relative entropy can be computed by means of a standard replica QFT calculation. We anticipate that this relative entropy contribution will be subleading to the gravitational contribution. We will see that the $S_{\rm rel}(\Phi)$ is exponentially small in regions $I$ and $I\nspc I$ and reaches a maximum of order $\sim \log 2$ in the neighborhood of the transition point $\tau=\tau_c$ between the two regions.

The replica QFT calculation for the relative entropy associated to an interval was studied in \cite{relqft}, following the classic work of Cardy and Calabrese \cite{cc}. The key idea of \cite{cc} is that the 
$n$-replica geometry is an $n$-fold branched cover $\Sigma_n$ of the original spacetime. The QFT path integral on $\Sigma_n$ involves the insertion of twist fields that create an angle excess equal to $2\pi n$  at the end-points of the entanglement cut. We expect to find this same geometric prescription in our set up. A schematic argument goes as follows.

Introduce the  $n$-th relative R\'enyi entropy via
\bea
\label{relrenyi}
S_{{\rm rel}, n}(\Phi|\!\smpc|\Psi) \is
\frac{1} {1\nspc-\nspc n}\log \la \smpc \Phi | \Delta_{\Psi|\Phi}^{1-n}  |\Phi \smpc\ra
\eea
The Araki formula is recovered upon taking the $n\to 1$ limit.
The right-hand side of \eqref{relrenyi} can be recast in a more practical form via a straightforward (though still slightly subtle) repeated application of the identity $\Delta_{\Phi}\Delta^{-1}_{\Psi|\Phi}= \Delta_{\Phi|\Psi}\Delta^{-1}_{\Psi}$, the explicit form of the Hawking pair state $\Phi$, the KMS condition, the definition of the modular time flow, and  the defining property of the relative modular operator  $\Delta_{\Phi|\Psi}$ that $\la \Psi| {\bf b} \Delta_{\Phi|\Psi}{\bf a} |\Psi\ra = \la \Phi| {\bf a}{\bf b} |\Phi\ra$ for any ${\bf a},{\bf b} \in {\cal A}$. This inductive procedure leads the following identity
\bea
\label{relrep}
{\la \Phi|  \spc \Delta_{\Psi|\Phi}^{1-n} \spc |\Phi\ra} 
\is \frac{1}{{Z}_{\beta,\tau}^{n}\!} \; \la \Psi| \spc \prod_{k=-1}^{2n-2} \!  {\phi}{}_{\nspc{\raisebox{-1.5pt}{\scriptsize $\tau \! - \! k   \frac \beta 2$} }}\spc
\, \bar{\phi}_{{\raisebox{-1.25pt}{\scriptsize $\tilde\tau\! -\! k \frac \beta 2$}}}\,
 |\Psi\ra
 \eea
 where the product over $k$  is (cyclicly) time ordered. We observe that, in contrast to the gravitational contribution, the operator insertions  in \eqref{relrep} are replicated, while the inverse temperature is not rescaled by a factor of $n$. The operator insertions, on the other hand, do span the full range of euclidean times of an $n$-fold replica geometry.

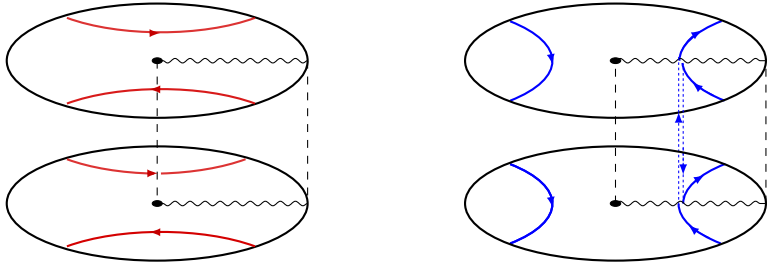
\begin{figure}
\begin{center}
\begin{tikzpicture}[xscale=1, yscale=.38, baseline={([yshift=0cm]current bounding box.center)}]);
\draw[yshift=2cm][thick] (4,0) arc (0:360:2);
\draw[yshift=2cm][thick, color={rgb:red,10; black,2},, opacity=.9] (.8,-1.5) arc (90+44:90-44:1.8);
\draw[yshift=2cm][thick, color={rgb:red,10; black,2}, opacity=.8] (.8,1.5) arc (270-44:270+44:1.8);
\draw[yshift=2cm][decorate, decoration={snake, segment length=2mm, amplitude=.3mm}] (4,0) -- (2,0);
\draw[yshift=2cm][xscale=.667,fill=black] (3,0) circle (0.1);
\draw[yshift=-3cm][thick] (4,0) arc (0:360:2);
\draw[yshift=2cm][dashed] (2,0) -- (2,-5);
\draw[yshift=2cm][dashed] (4,0) -- (4,-5);
\draw[yshift=-3cm][thick, color={rgb:red,10; black,2}, opacity=1] (.8,-1.5) arc (90+44:90-44:1.8);
\draw[yshift=-3cm][thick, color={rgb:red,10; black,2}, opacity=.8] (.8,1.55) arc (270-44:267:1.8);
\draw[yshift=-3cm][thick, color={rgb:red,10; black,2}, opacity=.8] (2.05,1.05) arc (274:270+44:1.8);
\draw[yshift=2cm][thin, color={rgb:red,10; black,3},latex-,opacity=1] (1.9,-1) arc (268:272:2);
\draw[yshift=-3cm][thin, color={rgb:red,10; black,3},latex-,opacity=1] (1.9,-1) arc (268:272:2);
\draw[yshift=2cm][thin, color={rgb:red,10; black,3},-latex,opacity=1] (1.9,.97) arc (268:272:2);
\draw[yshift=-3cm][thin, color={rgb:red,10; black,3},-latex,opacity=1] (1.87,1.06) arc (268:272:2);
\draw[yshift=-3cm][decorate, decoration={snake,segment length=2mm, amplitude=.3mm}] (4,0) -- (2,0);
\draw[yshift=-3cm][xscale=.667, fill=black] (3,0) circle (0.1);
\end{tikzpicture}~~~~~~~~~~~~~~~\begin{tikzpicture}[xscale=1, yscale=.38, rotate=90,baseline={([yshift=0cm]current bounding box.center)}]);
\draw[xshift=2cm][thick, color={rgb:blue,10; black,0},opacity=6] (.6,-1.45) arc (90+44:92:2);
\draw[xshift=2cm][color={rgb:blue,10; black,0},-latex,opacity=6] (.8,-1.30) arc (130:130-15:2);
\draw[xshift=2cm][thick, color={rgb:blue,10; black,0}] (2,-.85) arc (90:90-44:2);
\draw[xshift=2cm][thin, color={rgb:blue,10; black,0},-latex] (2.2,-.86) arc (85:90-32:1.95);
\draw[xshift=2cm][thick, color={rgb:blue,10; black,0},opacity=4] (.6,1.4) arc (270-44:270+44:2);
\draw[xshift=2cm][decorate, decoration={snake,segment length=2mm, amplitude=.3mm}](2,0) -- (2,-2);
\draw[xshift=2cm][thin, color={rgb:blue,10; black,0},-latex,opacity=4] (2,.85) arc (278:274:2);
\draw[xshift=2cm][yscale=.69,fill=black] (2,0) circle (0.1);
\draw[xshift=2cm][thick] (4,0) arc (0:360:2);
\draw[xshift=-3cm][dashed] (2,0) -- (7,0);
\draw[xshift=-3cm][thin,color={rgb:blue,10; black,0}, dashed,dash pattern=on 1pt off 1pt,-latex,opacity=2] (1.95,-.84) -- (5.2,-.84);
\draw[xshift=-3cm][thin,color={rgb:blue,10; black,0}, dashed,dash pattern=on 1pt off 1pt,opacity=2] (5,-.84) -- (7.15,-.84);
\draw[xshift=-3cm][color={rgb:blue,10; black,0},-latex,opacity=1] (.8,-1.25) arc (130:130-15:2);
\draw[xshift=-3cm][thin,color={rgb:blue,10; black,0}, dashed,dash pattern=on 1pt off 1pt, latex-,opacity=6] (3,-.9) -- (6.9,-.9);
\draw[xshift=-3cm][thin,color={rgb:blue,10; black,0}, dashed,dash pattern=on 1pt off 1pt, opacity=6] (2,-.9) -- (4,-.9);
\draw[xshift=-3cm][dashed] (2,-2) -- (7,-2);
\draw[xshift=-3cm][thick, color={rgb:blue,10; black,0}] (.6,-1.4) arc (90+44:90:2);
\draw[xshift=-3cm][thick, color={rgb:blue,10; black,0},opacity=6] (2,-.9) arc (90:90-44:2);
\draw[xshift=-3cm][thin,color={rgb:blue,10; black,0}, -latex, opacity=6] (2.62,-1) arc (90-20:90-32:2);
\draw[xshift=-3cm][thick, color={rgb:blue,10; black,0},opacity=4] (.6,1.4) arc (270-44:270+44:2);
\draw[xshift=-3cm][thin, color={rgb:blue,10; black,0},-latex,opacity=4] (2,.85) arc (278:274:2);
\draw[xshift=-3cm][thick, color={rgb:blue,10; black,0},opacity=4] (.6,1.4) arc (270-44:270+44:2);
\draw[xshift=-3cm][decorate, decoration={snake, segment length=2mm, amplitude=.3mm}](2,0) -- (2,-2);
\draw[xshift=-3cm][yscale=.69,fill=black] (2,0) circle (0.1);
\draw[xshift=-3cm][thick] (4,0) arc (0:360:2);
\end{tikzpicture}
\end{center}
\vspace{-2mm}
\caption{The replica geometry for the relative entropy, here shown for $n=2$, is a branched cover of the euclidean black hole spacetime with total excess angle $2\pi n$ located at the horizon.  The boundary-to-boundary propagators equal the thermal two-point function at inverse temperature~$\beta$. Depending on $\tau$, either the left or right Wick contraction dominates.}
\vspace{-2mm}
\end{figure}

Combining the above two observations reproduces the Cardy-Calabrese prescription that expresses the expectation value \eqref{relrep} as the QFT path integral over an $n$-sheeted branched cover $\Sigma_n$ of the euclidean black hole geometry (or `thermal disk'), 
as shown in figure 4 for $n=2$
 \bea
 \label{relrept}
{\la \Phi|  \spc \Delta_{\Psi|\Phi}^{1-n} \spc |\Phi\ra}   \is\frac{1}{{Z}_{\beta,\tau}^{n}\!} \; \Bigl\langle \prod_{k=-1}^{2n-2} \!\!  {\phi}{}_{\nspc{\raisebox{-1.5pt}{\scriptsize $\tau  \nspc\! - \! k \nspc  \frac \beta 2$} }}\spc
\, \bar{\phi}_{{\raisebox{-1.25pt}{\scriptsize $\tilde\tau\!\nspc -\! k \frac \beta 2$}}}
\Bigr\rangle_{\Sigma_n}
\eea
The black hole horizon is now the location of a conical singularity with a total angle excess equal to $2\pi (n-1)$. Away from the horizon, the bulk geometry locally looks identical to the black hole with $n=1$. Hence the boundary-to-boundary propagator equals the thermal two-point function at inverse temperature~$\beta$. The geometric path-integral prescription \eqref{relrept} for the replica correlator thus reproduces the operator expectation value \eqref{relrep}.

The next steps are routine: we write the replica correlator as a sum over Wick contractions and only keep the dominant term with the shortest nearest neighbor contractions. Depending on the value of $\tau$, either the left or right Wick contraction shown in figure 4 dominates. Generalizing to arbitrary $n$, we find 
\bea
\label{relrepf} \Bigl\langle\, \prod_{k=-1}^{2n\nspc -\nspc 2} \!\!  {\phi}{}_{\nspc{\raisebox{-1.5pt}{\scriptsize $\tau  \!\! - \!\nspc k \nspc  \frac \beta 2$} }}
\, \bar{\phi}_{{\raisebox{-1.25pt}{\scriptsize $\tilde\tau\nspc - \nspc k \frac \beta 2$}}}
\Bigr\rangle_{\!\Sigma_n}\!\!\!\! \is  \!  G_{\beta} (\spc y \spc)^{2n}  \qquad \quad y=\min(2\tau,\tau\nspc-\nspc \tilde{\tau})
\eea
The contributions from other Wick contractions are exponentially suppressed. 

Plugging the leading order result \eqref{relrepf} into \eqref{relrep} and using the result \eqref{twocases} for $Z_{\beta,\tau}$, we find that the numerator and denominator on the right-hand side cancel   ${\la \Phi|  \spc \Delta_{\Psi|\Phi}^{1-n} \spc |\Phi\ra}  \simeq  1,$  both in the near horizon and in the far away regions. This confirms our expectation that the relative R\'enyi entropy vanishes 
to leading order in the semi-classical approximation. 
The relative entropy $S_{\rm rel}(\Phi|\!\smpc|\Psi)$ is the only measure of quantum information transfer that is identifiable from the perspective of QFT in curved spacetime. The fact that it vanishes at leading order lies at the root of the black hole information paradox in its original form.

The leading non-zero contribution to the relative entropy is of order unity and concentrated in the transition region near $\tau = \tau_c = \beta/8$.
  Taking into account that the two types of nearest-neighbor Wick contractions are of comparable magnitude in the transition region, we find that (up to higher order terms)
\bea
\label{hpairs}
S_{\rm rel} (\Phi|\!\smpc|\Psi)  \,=\, &-&\!\!\!  p_0 \log p_0 - p_1\log p_1 \qquad \ 
p_1 = 1-p_0 = \frac{{G}_\beta^2}{{G_\beta^2} + {\tilde{G}_\beta^2}},
\eea
with $G_\beta = G_\beta(\spc{x}\spc)$ and $\tilde{G}_\beta = G_\beta(\spc\tilde{x}\spc)$. 
Corrections to \eqref{hpairs} are exponentially suppressed for $h\gg 1$. In view of our earlier discussion at the end of section 2.3, the formula \eqref{hpairs} can be interpreted as the entropy of the `Hawking qubit'  that keeps track of the occupation number of the exterior Hawking particle
\bea
\label{hq}
\rho_{\rm {}_{Hawking\; qubit}} \is p_0 |0\ra \la 0 | + p_1 |1\ra \la 1| 
\eea
where $|0\ra$ and $|1\ra$ are a short-hand notation for whether the Hawking pair is present or not.  The formula \eqref{hpairs} gives an accurate description of the relative entropy in the transition region $\tau\simeq \tau_c$ and correctly captures the fact that $p_1$ decays exponentially in region $I$ and $p_0$ decays exponentially in region $I\nspc I$. 

The relative entropy has a maximum $S_{\rm rel} = \log 2$ at the transition point $\tau= \tau_c$ where the 0 and 1 occupancy state have equal probability. Far away from $\tau=\tau_c$, either the 0 occupancy or the 1 occupancy state dominates, and the relative entropy goes to zero in both regions.
Hence the Hawking qubit \eqref{hq} is not  capable for extracting quantum information from the black hole. Instead, the black hole quantum information is carried out via microscopic (ultra-soft) degrees of freedom, that can not (in any obvious way) be captured via a standard local QFT description.

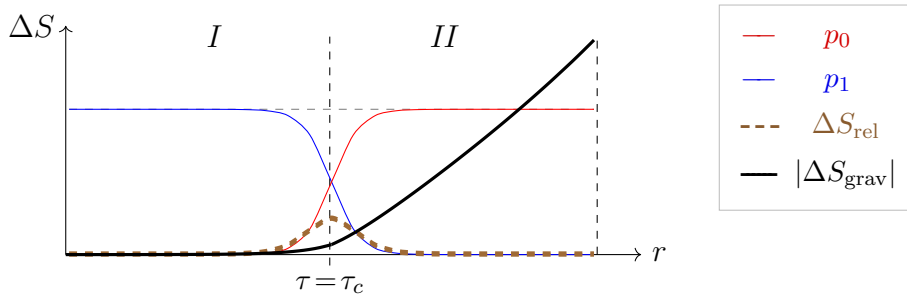
\begin{figure}[t]
\begin{center}
~~~~~~~~~\begin{tikzpicture}[xscale =4.5,yscale = 2.9,domain=0.01:pi/2-0.01]
  \draw[->] (0, 0) -- (1.7, 0) node[right] {$r$~~~~${}$};
  \draw[->] (0, 0) -- (0, 1.05) node[left] {${\Delta S}$};
  \draw[dashed] (.78, -.05) node[below] {\small $\tau\!=\!\tau_c$} -- (.78,1);
  \draw[dashed] (1.57, 0)-- (1.57,1); 
  \matrix [draw,lightgray,below right] at (current bounding box.north east) { \\
   \node [color={rgb:red,5; black,1}]  {\small $-\!\!\!- \;\;\; \;\;p_0$}; \\
   \node [color={rgb:blue,10; black,1}]  {\small  $-\!\!\!- \;\;\; \;\;p_1$}; \\
   \node [color={rgb:brown,10; black,4}]  {\small  {\bf -\;\!-\;\!-}  $\; \; \, \Delta S_{\rm rel}$}; \\
  \node [color={rgb:black,5; black,5}] {\small  \!\! {\bf -\!-\!-\!-\!-\!-\,} $|\Delta S_{\rm grav}|$};\\
};
\draw (1.12,1) node{$I\nspc I$};
\draw (.44,1) node{$I$};
    \draw[thin,color=gray,dashed]   plot  [smooth] (\x,{2/3})  ;
    \draw[color=blue]   plot  [smooth] (\x,{(-tanh(12*(\x-pi/4))+1)/3})  ;
    \draw[color=red]   plot  [smooth] (\x,{(tanh(12*(\x-pi/4))+1)/3})  ;
    \draw[very thick,color={rgb:brown,10; black,2},dashed]   plot  [smooth] (\x,{((-(tanh(12*(\x-pi/4))+1)/2)* log2((tanh(12*(\x-pi/4))+1)/2+.01)-((-tanh(12*(\x-pi/4))+1)/2)* log2((-tanh(12*(\x-pi/4))+1)/2+.01))/6}) ;
    \draw[very thick,color={rgb:brown,10; black,2},dashed]   plot  [smooth] (\x,{((-(tanh(12*(\x-pi/4))+1)/2)* log2((tanh(12*(\x-pi/4))+1)/2+.01)-((-tanh(12*(\x-pi/4))+1)/2)* log2((-tanh(12*(\x-pi/4))+1)/2+.01))/6+.01}) ;
    \draw[very thick,color={rgb:black,5; black,5}, domain=pi/4:pi/2-.01]  plot [smooth] (\x,{-(\x*cot(\x  r) - (pi/2-\x)*tan(\x r)) + 0.00002*(exp(10.2*(pi/2-\x))-1)}) ;
    \draw[very thick,color={rgb:black,5; black,5},  domain=0:pi/4]   plot [smooth] (\x,{0+0.00002*(exp(10.2*\x)-1)}) ;
\end{tikzpicture}
\vspace{-5mm}
\end{center}
\caption{Graph of the gravitational contribution $\Delta S_{grav}$ (solid black) and the relative entropy $\Delta S_{\rm rel}$ (dashed brown) as a function of the euclidean time $\tau$. The red and blue lines indicate the probability $p_0$ and  $p_1$ that the particle number of the exterior state is $0$ or $1$.}
\vspace{-2mm}
\end{figure}

\section{Conclusion}
\vspace{-1mm}

We introduced a new probe designed to keep track of the quantum information content of a Hawking pair as a function of its location. The probe is defined as the von Neumann entropy of the state $\Phi$  in \eqref{hpair} obtained by acting with a local single trace operator $\phi_\tau$ and its mirror $\bar{\phi}_{\tilde\tau}$ on a microcanonical state $\Psi$ of energy $E$ given in \eqref{micropure}. We computed the entropy difference $\Delta S = S(\rho_\Phi)\nspc - \nspc S(\rho_\Psi)$ via thermodynamics, via a replica method, by using JT gravity, and by applying the operator algebra treatment introduced in \cite{wittenetal}. Here we collect some comments on the physical interpretation and implication of our result.

\smallskip
\smallskip

{\it Interpretation of $\Delta S$}. The treatment of \cite{wittenetal} allows one to separate the entropy difference $\Delta S$ into a bulk QFT contribution $\Delta S_{\rm rel}$ and a gravitational contribution $\Delta S_{\rm grav}$. The former vanishes everywhere, except near the critical radius $\tau=\tau_c$ where the probability $p_0$ that $\Phi$ contains no extra Hawking pair is comparable to the probability $p_1$ that it contains one extra Hawking pair. The latter keeps track of the microscopic quantum information extracted from the black hole and dominates over the former in the regime of interest. As shown in figure 5,  the microcanonical gravitational entropy difference vanishes in the near horizon regime $I$, where $p_0$ is close to 1, and rises linearly in the far away region $I\nspc I$, where $p_1$ is close to~1. This behavior of $\Delta S_{\rm grav}$ indicates that the amount of microscopic quantum information that the Hawking pair extracts from the black hole is vanishingly small in the near horizon region   $\tau>\tau_c$  to the left the critical point $\tau_c=\beta/8$ and then linearly grows to its maximal value $|\Delta S_{\rm grav}|= 4h$ near the boundary at $\tau=0$. 

The fact that $\rho_\Phi$ has smaller von Neumann entropy than the Hawking state $\rho_\Psi$ may seem a bit surprising since the extra pair would seem to add rather than subtract entropy. From a microscopic perspective, however, it is clear that specifying the presence of the extra Hawking pair at some location reduces the entropy, for two separate reasons.  

 An AdS black hole does not evaporate on its own. To let it evaporate, we must introduce a measuring apparatus $A$ that can detect and absorb external Hawking particles. Let $\Phi_i = \int\! d\tau\, f_i(\tau) \Phi_{\tau}$ denote an orthonormal basis of wave functions, and $| i \ra_{\! A}$ denote the state of the apparatus $A$ after it has measured the Hawking particle in state $\Phi_i$. We can model the unitary time step from detecting no particle to detecting one particle via 
\bea
U \bigl(\Psi \otimes |0\ra_{{\!}A}\bigr) \is \sum_{i} \alpha_i \spc \Phi_i \otimes |i \ra_A
\eea
where $U$ is the time evolution operator of the external Hawking state interacting with the measuring apparatus. Unitary time evolution leaves the total external entropy invariant
\bea
\Delta S_{\rm tot} \is \sum_i \spc |\alpha_i|^2 \spc \Delta S_i  \spc  -\spc \sum_i   |\alpha_i|^2 \log  |\alpha_i|^2  = 0
\eea
with $\Delta S_i = S(\rho_{\Phi_i})\nspc -\nspc S(\rho_\Psi)$. The second contribution is positive, the first is negative.

\medskip

\noindent{\it Microscopic mechanism}. What is the microscopic entropy extraction mechanism? A key property of the CFT relative to the effective bulk QFT description of the Hawking pair is that the local operators $\phi_\tau$ and $\bar\phi_{\tilde\tau}$ act as linear maps on the microscopic Hilbert space with energy $E$ and entropy $S(E)$.  As indicated in figure 6, this allows for the possibility that the Hawking pair produces an intermediate channel with either $\; I:$~larger energy $E'>E$ and entropy $S(E')>S(E)$, or $I\nspc I:$~smaller energy $E'<E$ and entropy $S(E')<S(E)$. In~the latter case, the Hawking pair creates an information bottle neck that filters out some of the quantum information contained in the initial black hole. Interestingly, the transition point between the two entropic regimes $I$ and $I\nspc I$ coincides with the point where the blue QFT Wick contraction in figure \eqref{channels} becomes dominant over the red contraction. The linkage between the entropic transition and the geometric transition is made manifest by the random matrix description of the $\phi_\tau$ correlation functions introduced above equation \eqref{cwick}. This correspondence motivates the general holographic conjecture that {\it the dominant Wick contraction in any bulk effective field theory calculation takes place along the channel with the largest number of intermediate microscopic states.}

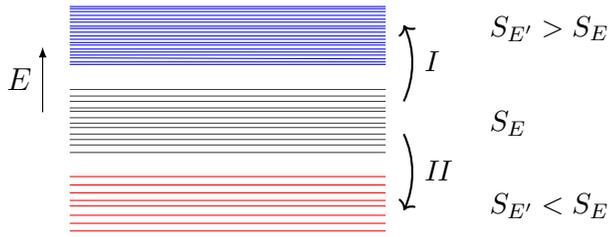
\begin{figure}[t]
\begin{center}
${}$~~~~~~~~~~\begin{tikzpicture}[xscale=1.2,yscale=1.45]
\draw[thin,\blueone] (0,0)-- (3.5,0);
\draw[thin,\blueone] (0,0.1)--(3.5,0.1)node[right]{$\raisebox{-1pt}{\textcolor{black}{\qquad \quad \small $S_E$}}$};
\draw[thin,\blueone] (0,0.06)-- (3.5,0.06);
\draw[thin,\blueone] (0,0.15)-- (3.5,0.15);
\draw[thin,\blueone] (0,0.209)-- (3.5,0.209);
\draw[thin,\blueone] (0,0.3)-- (3.5,0.3);
\draw[thin,\blueone] (0,0.24)-- (3.5,0.24);
\draw[thin,\blueone] (0,0.349)-- (3.5,0.349);
\draw[thin,\blueone] (0,0.409)-- (3.5,0.409);
\draw[thin,\blueone] (0,-0.1)-- (3.5,-0.1);
\draw[thin,\blueone] (0,-0.169)-- (3.5,-0.169);
\draw[thin,\blueone] (0,-0.05)-- (3.5,-0.05);
\draw[thin,\bluetwo] (0,1.17)-- (3.5,1.17);
\draw[thin,\bluetwo] (0,0.725)-- (3.5,0.725);
\draw[thin,\bluetwo] (0,0.662)-- (3.5,0.662);
\draw[thin,\bluetwo] (0,0.78)-- (3.5,0.78);
\draw[thin,\bluetwo] (0,0.639)-- (3.5,0.639);
\draw[thin,\bluetwo] (0,0.817)-- (3.5,0.817);
\draw[thin,\bluetwo] (0,0.699)-- (3.5,0.690);
\draw[thin,\bluetwo] (0,0.753)-- (3.5,0.753);
\draw[thin,\bluetwo] (0,0.9)-- (3.5,0.9);
\draw[thin,\bluetwo] (0,0.87)-- (3.5,0.87);
\draw[thin,\bluetwo] (0,0.969)--(3.5,0.969)node[right]{$\raisebox{-1pt}{\textcolor{black}{\qquad \quad \small$S_{E'}>S_E$}}$};;
\draw[thin,\bluetwo] (0,0.935)-- (3.5,0.935);
\draw[thin,\bluetwo] (0,0.84)-- (3.5,0.84);
\draw[thin,\bluetwo] (0,0.99)-- (3.5,0.99);
\draw[thin,\bluetwo] (0,1.027)-- (3.5,1.027);
\draw[thin,\bluetwo] (0,1.054)-- (3.5,1.054);
\draw[thin,\bluetwo] (0,1.089)-- (3.5,1.089);
\draw[thin,\bluetwo] (0,1.15)-- (3.5,1.15);
\draw[thin,\bluetwo] (0,1.124)-- (3.5,1.124);
\draw[thin,\bluethree] (0,-0.609)-- (3.5,-0.609);
\draw[thin,\bluethree] (0,-0.539)-- (3.5,-0.539);
\draw[thin,\bluethree] (0,-0.467)-- (3.5,-0.467);
\draw[thin,\bluethree] (0,-0.745)-- (3.5,-0.745);
\draw[thin,\bluethree] (0,-0.66)--(3.5,-0.66)node[right]{$\raisebox{-1pt}{\textcolor{black}{\qquad \quad \small$S_{E'}<S_E$}}$};
\draw[thin,\bluethree] (0,-0.819)-- (3.5,-0.819);
\draw[thin,\bluethree] (0,-0.889)-- (3.5,-0.889);
\draw[thin,\bluethree] (0,-0.95)-- (3.5,-0.95);
\draw[thin,\bluethree] (0,-0.39)-- (3.5,-0.39);
\draw[thin,-latex] (-.3,.2) -- (-.3,.5)node[left] {$E$} -- (-.3,.8);
\draw[thick,->] (3.7,0.3) arc (330:390:.7)node[above,right]{$\raisebox{-35pt}{\textcolor{black}{\small\ $I$}}$};  
\draw[thick,<-] (3.7,-0.7) arc (330:390:.7)node[below,right]{$\raisebox{-35pt}{\textcolor{black}{\small\ $I\nspc I$}}$}; 
\end{tikzpicture}
\vspace{-1mm}
\end{center}
\caption{Microscopic energy spectrum before and after Hawking emission. The emission process maps the Hilbert subspace ${\cal H}_E$ with entropy $S(E)$ to a subspace with ${\cal H}_{E'}$ with entropy $S(E')$. If  $S_{E'}>S_E$ the map is invertible, if $S_{E'}>S_E$  it is non-invertible.} 
\vspace{-1mm}
\end{figure}

\medskip

\noindent{\it Mini Page curve}.  As noted in section 5.1, in the canonical ensemble with fixed temperature, the entropy of the Hawking pair state follows a Page-like curve shown in figure 3. Indeed, the black hole information paradox has the following differential formulation. 

Consider a Hawking particle that escapes a black hole held at fixed temperature.  The presence of the extra particle initially contributes an entropy of order $h\gg 1$, given in equation~\eqref{deltascan}.  The physical source of the extra entropy is the additional microscopic entanglement between the external state with the particle and the black hole. This extra entanglement is not just that of the `Hawking qubit' itself, but includes entanglement of microscopic quantum fluctuations. After the particle travels onward and escapes through the AdS boundary, the black hole returns to its original thermal state with inverse temperature $\beta$. Hence the final and initial entropy are equal. So, somewhere along its way from the horizon to the AdS boundary, the microscopic state of the Hawking particle must lose its entanglement with the black hole interior in exchange for  entanglement with the external heat bath. This entanglement swap seems impossible without violating locality.

This differential formulation of the information paradox  and the mini-Page curve are as fundamental as the global formulation of the paradox and the  global Page-curve. A complete description of a Hawking pair emission should explain how it extracts $\Delta S = \beta \Delta E$ amount of quantum information. The mini-Page curve encapsulates this information transfer. Stringing together many successive Hawking pair emissions by an initial pure state black hole then builts up the macroscopic Page curve of the full evaporation process. The circumstances and mechanism by which Hawking pairs extract quantum information from a black hole do not depend on whether the black hole is in a pure state or a mixed state. Each individual Hawking pair emission has the same microscopic and macroscopic description, regardless of the time epoch relative to the Page-time.

\section*{Acknowledgements}
\vspace{-2mm}

We thank Ahmed Almheiri, Scott Collier, Chelsea Ding, Akash Goel, Kanato Goto, Clifford Johnson, Monica Kang, Henry Lin, Bowei Liu,  Juan Maldacena, Henry Maxfield, Geoff Penington,  Erik Verlinde, Edward Witten, and Mengyang Zhang for valuable discussions and comments. This research is supported by NSF grant PHY-2209997.

\appendix

\section{Backreaction in JT gravity}
\vspace{-1.5mm}
In this Appendix we summarize the calculation of the entropy of the Hawking pair state in JT gravity \cite{JT}-\cite{Zhenbin}. The treatment directly follows the earlier work \cite{GLTV} on the entropy of partially entangled states. The calculation relies purely on the classical backreaction of the Hawking pair on the black hole geometry, or rather, on the value of the dilaton at the horizon, which in JT gravity accounts for the Bekenstein-Hawking entropy of the black hole.

The dynamics of euclidean JT gravity can be reformulated as that of a charge boundary particle moving in a constant electric field on the hyperbolic plane \cite{KitaevSuh}-\cite{Zhenbin}. The trajectory of the boundary particle indicates the edge of the nearly AdS2 space-time. Without matter, the trajectory takes the form of a thermal circle with length equal to the inverse temperature $\beta$. The energy of the boundary particle is conveniently parametrized by means of the continuous spin quantum number of $SL(2,\mathbb{R})$, the isometry group of AdS2.  

As shown in figure 7, the insertion of the Hawking pair introduces two boundary-to-boundary propagators, or semi-classically, two boundary-anchored bulk geodesics that backreact on the boundary trajectory. As before, there are two Wick contractions. The operator insertions split the thermal circle into four segments. Due to the reflection symmetry, the segments on opposite sides have identical quantum numbers, labeled by the continuous $SL(2,\mathbb{R})$ spins denoted by $p$ and $q$. The energy is proportional to the $SL(2,\mathbb{R})$ Casimir and entropy is determined by the Plancherel measure on the space of representations
\bea
E = \spc \frac{q^2}{2C}, \quad & & \quad S(E)\spc =\spc 2\pi q \spc = \spc 2\pi \sqrt{2CE} \nn\\[-2mm]\\[-2mm]\nn
E' =\spc \frac{p^2}{2C}, \quad & &\quad S(E')\spc =\spc 2\pi p \spc = \spc 2\pi \sqrt{2CE'} 
\eea
Here $C$ is the coupling constant of the Schwarzian quantum mechanics that describes the motion of the boundary particle. $S(E)$ is the energy dependent contribution to the entropy; in general, there is also a constant contribution $S_{\rm tot}(E) = S_0 + S(E)$. This formula for the entropy matches with the value of the dilaton at the horizon of a JT gravity black hole with mass $M=E$.

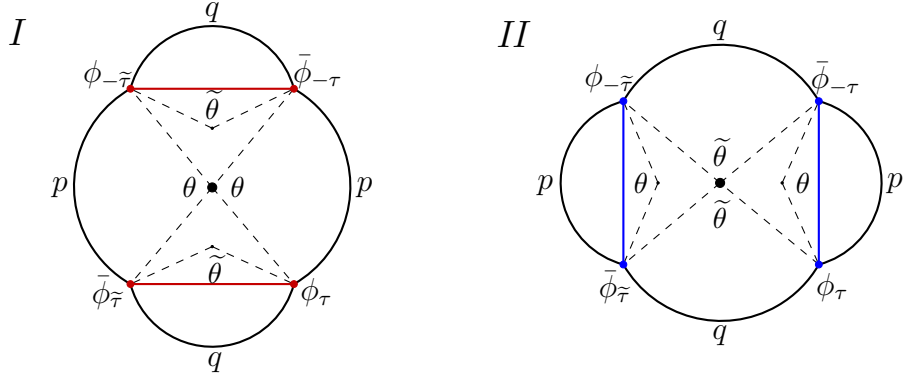
\begin{figure}[t]
\vspace{-0.6cm}
\begin{center}
\raisebox{-.36cm}{$\begin{tikzpicture}[scale=.75,rotate = 90]
\draw (1,2) node {\large $I$};
\draw[thick] (0,0) coordinate (a) arc (30:150:2) coordinate (e); 
\draw[thick] (e) arc (105:255:1.5) coordinate (f); 
\draw[thick] (f) arc (210:330:2) coordinate (g); 
\draw[thick] (0,0) arc (75:-75:1.5) coordinate (d); 
\draw (a)+(0.25,0.4) node {${\phi}_{-\tilde\tau}$};
\draw (e)+(-0.15,0.4) node {$\bar{\phi}_{\tilde{\tau}}$};
\draw (f)+(-0.15,-0.4) node {${\phi}_\tau$};
\draw (d)+(0.3,-0.4 25) node {$\bar\phi_{-\tau}$};
\draw[thick,color={rgb:red,10; black,3}] (e) -- (f);
\draw[thick,color={rgb:red,10; black,3}] (a) -- (d);
\draw[fill=black] (-1.75,-1.45) coordinate (b) circle (0.08);
\draw[fill=black] (-0.7,-1.45) coordinate (c) circle (0.02);
\draw[fill=black] (-2.8,-1.45) coordinate (o) circle (0.02);
\draw[dashed] (b) -- (a);
\draw[dashed] (d) -- (b);
\draw[dashed] (b) -- (e);
\draw[dashed] (b) -- (f);
\draw[dashed] (o) -- (e);
\draw[dashed] (o) -- (f);
\draw[dashed] (c) -- (d);
\draw[dashed] (c) -- (a);
\draw (-0.28,-1.45) node {\footnotesize $\tilde\theta$};
\draw (-3.2,-1.5) node {\footnotesize $\tilde\theta$};
\draw (-1.75,-1.05) node {\footnotesize $\theta$};
\draw (-1.75,-1.9) node {\footnotesize $\theta$};
\draw (1.35,-1.45) node {$q$};
\draw (-4.85,-1.5) node {$q$};
\draw (-1.75,1.25) node {$p$};
\draw (-1.75,-4.15) node {$p$};
\draw[color={rgb:red,10; black,3}, fill={rgb:red,10; black,3}] (a) circle (0.06);
\draw[color={rgb:red,10; black,3}, fill={rgb:red,10; black,3}] (d) circle (0.06);
\draw[color={rgb:red,10; black,3}, fill={rgb:red,10; black,3}] (e) circle (0.06);
\draw[color={rgb:red,10; black,3}, fill={rgb:red,10; black,3}] (f) circle (0.06);
\end{tikzpicture}$}~~~~~~~~~~{\begin{tikzpicture}[scale=.75]
\draw (-5.4,1.2) node {\large $I\nspc I$};
\draw[thick] (0,0) coordinate (a) arc (30:150:2) coordinate (e); 
\draw[thick] (e) arc (105:255:1.5) coordinate (f); 
\draw[thick] (f) arc (210:330:2) coordinate (g); 
\draw[thick] (0,0) arc (75:-75:1.5) coordinate (d); 
\draw (a)+(0.3,0.4) node {$\bar\phi_{-\tau}$};
\draw (e)+(-0.25,0.4) node {${\phi}_{-\tilde\tau}$};
\draw (f)+(-0.15,-0.4) node {$\bar{\phi}_{\tilde{\tau}}$};
\draw (d)+(0.24,-0.395) node {${\phi}_\tau$};
\draw[thick,color={rgb:blue,10; black,0}] (e) -- (f);
\draw[thick,color={rgb:blue,10; black,0}] (a) -- (d);
\draw[fill=black] (-1.75,-1.45) coordinate (b) circle (0.08);
\draw[fill=black] (-0.65,-1.45) coordinate (c) circle (0.02);
\draw[fill=black] (-2.85,-1.45) coordinate (o) circle (0.02);
\draw[dashed] (b) -- (a);
\draw[dashed] (d) -- (b);
\draw[dashed] (b) -- (e);
\draw[dashed] (b) -- (f);
\draw[dashed] (o) -- (e);
\draw[dashed] (o) -- (f);
\draw[dashed] (c) -- (d);
\draw[dashed] (c) -- (a);
\draw (-0.28,-1.45) node {\footnotesize $\theta$};
\draw (-3.15,-1.45) node {\footnotesize $\theta$};
\draw (-1.75,-0.9) node {\footnotesize $\tilde\theta$};
\draw (-1.75,-2) node {\footnotesize $\tilde\theta$};
\draw (1.35,-1.45) node {$p$};
\draw (-4.85,-1.5) node {$p$};
\draw (-1.75,1.25) node {$q$};
\draw (-1.75,-4.15) node {$q$};
\draw[color={rgb:blue,10; black,0}, fill={rgb:blue,10; black,0}] (a) circle (0.06);
\draw[color={rgb:blue,10; black,0}, fill={rgb:blue,10; black,0}] (d) circle (0.06);
\draw[color={rgb:blue,10; black,0}, fill={rgb:blue,10; black,0}] (e) circle (0.06);
\draw[color={rgb:blue,10; black,0}, fill={rgb:blue,10; black,0}] (f) circle (0.06);
\end{tikzpicture}}
\end{center}
\vspace{-0.35cm}
\caption{\small The boundary curve with two pairwise operator insertions. The left figure is the dominant saddle point in region $I$, the right figure is the dominant sadde point in region $I\nspc I$. The red and blue lines indicate the bulk propagators and the angles $\theta$ and $\tilde{\theta}$ parametrize the backreaction of the bulk particles on the boundary curve.}
\label{fig:curve2}
\end{figure}

Correlation functions in JT gravity with multiple boundary-to-boundary propagators are exactly known. As shown in \cite{GLTV}, they can be cast into the form of an integral expression that, in the semi-classical limit, reduces to a saddle-point formula. For the four-point functions, the saddle point formula takes the following form
\bea
Z_{\beta,\tau} = \exp\bigl({-F_{\beta,\tau}}\bigr) , \quad& &  \quad F_{\beta,\tau} = \; \min_{p,q,\theta,\tilde{\theta}}  \  I_{\beta,\tau}(p,q,\theta,\tilde\theta)\\[-8mm]\nn
\eea
where\\[-8mm]
\bea
\label{integrand}
 I_{\beta,\tau}(p,q,\tilde\theta,\theta)\!\!\is \!\! (2\theta\nspc -\nspc 2\pi n) \spc p \spc + \,(2\tilde\theta\nspc -\nspc 2\pi\smpc \tilde{n})\spc q \spc +
\frac{p^2}{C} \spc {x}\, + \, \frac{q^2}{C} \spc  \tilde{x} \spc +\spc 2h \log\Bigl( \cos \frac{\theta} 2\! + \nspc\cos \frac {\tilde\theta} 2\,\Bigr)^2  \nn
\eea
where $\tilde{x} = x-\beta/2$. The variables $p$ and $q$ label the intermediate energy eigenstates. The numbers $n$ and $\tilde{n}$ are multiplicities
\bea
 I\spc :\  \ \ \   n=2,\tilde{n} = 1 \qquad & & \qquad 
 I\nspc I :\ \ \ \  n=1,\tilde{n} =2
\eea
that account for the fact that the energy of boundary segments that are connected through the bulk are constrained to be identical, whereas the intermediate energies in boundary segments that are separated by bulk propagators are independently summed over. The function $I_{\beta,\tau}(p,q,\tilde\theta,\theta)$ and the auxiliary variables $\theta$ and $\tilde{\theta}$ can be identified with the action and the geometric angles that specify the shape of the boundary trajectory, as indicated in figure 7. The saddle point equations read
\bea
\frac {\theta}  2 \! \is \!   \arctan \Bigl(\frac{p\nspc+\nspc q}{h} \Bigr)-   \arctan \Bigl(\frac{p\nspc-\nspc q}{h}\Bigr) \spc =\spc  \frac \pi {\raisebox{1pt}{$n$}} \spc -\spc \frac {p}C\spc {x} \ \nn\\[-2mm]\\[-2mm]\nn
\frac {\tilde\theta} 2\! \is\!   \arctan \Bigl(\frac{p\nspc+\nspc q}{h}\Bigr) +   \arctan\Bigl( \frac{p\nspc-\nspc q}{h}\Bigr) =  \spc \frac \pi {\raisebox{0pt}{$\tilde{n}$}}\spc-\spc \frac{q}C\spc \tilde{x}\ 
\eea
These are four equations with four unknowns. The general solution to these equations can not be written in elementary function. In the probe limit, where the back reaction due to the bulk propagators can be treated in the linearized approximation, one finds 
\bea S(E) \, = \, 2\pi q   \is S(E_\beta) + {4 h} \Bigl(1- \frac{\pi \tilde{x}}{\beta} \, {\cot} \Bigl(\spc \frac{\pi \tilde{x}}{\beta} \spc \Bigr)\Bigr)\qquad I\nn \\[-1.75mm]\\[-1.75mm]
S(E')\, = \, 2\pi p \is S(E_\beta) + {4 h} \Bigl(1- \frac{\pi x}{\beta} \, {\cot} \Bigl(\spc \frac{\pi x}{\beta} \spc \Bigr)\Bigr)\qquad I\nspc I\nn
\eea
This reproduces the result for the entropy obtained in section 4 via other methods, and confirms the interpretation of the entropy difference of the Hawking pair as the change in the Bekenstein-Hawking entropy due to gravitational backreaction of the pair.

\addtolength\baselineskip{-.5mm}

\end{document}